\documentclass[english]{article}
\usepackage[T1]{fontenc}
\usepackage[latin9]{inputenc}
\usepackage[english]{babel}
\usepackage{geometry}
\usepackage{setspace}
\usepackage{float}
\usepackage{graphicx}
\usepackage{chngcntr}
\usepackage[colorlinks=true, allcolors=black]{hyperref}

\usepackage{amssymb, amsfonts, amsmath}
\usepackage{bm}

\usepackage{adjustbox}
\usepackage{caption}            
\usepackage{rotating}           

\usepackage{booktabs}  
\usepackage{threeparttable}    
\usepackage{dcolumn}    
    \newcolumntype{d}[1]{D{.}{.}{#1}}

\usepackage[
    backend=biber,
    style=apa,
  ]{biblatex} 
\makeatletter

\let\SF@@footnote\footnote
\def\footnote{\ifx\protect\@typeset@protect
    \expandafter\SF@@footnote
  \else
    \expandafter\SF@gobble@opt
  \fi
}
\expandafter\def\csname SF@gobble@opt \endcsname{\@ifnextchar[
  \SF@gobble@twobracket
  \@gobble
}
\edef\SF@gobble@opt{\noexpand\protect
  \expandafter\noexpand\csname SF@gobble@opt \endcsname}
\def\SF@gobble@twobracket[#1]#2{}

\makeatother

\begin{document}
\title{Do Cure Violence Programs Reduce Gun Violence? Evidence from New York City \thanks{Avram: NYC Council Data Team, ravram@council.nyc.gov. Koepcke: NYC Council Data Team, ekoepcke@council.nyc.gov. Moussawi: NYC Council Data Team, amoussawi@council.nyc.gov. Nu\~nez: NYC Council Data Team, menunez@council.nyc.gov. The authors are grateful to Rose Martinez for her invaluable contributions to this paper. The authors thank Owen Kotowski for providing NYPD funding and staffing data and CMS budget data. The authors attest that, aside from the authors, this research was not conducted at the behest of any employee/member of the NYC Council and no other employee/member of the NYC Council had editorial control over this paper. Any opinions, findings, and conclusions or recommendations expressed in this material are those of the authors and do not necessarily reflect the views of the NYC Council. All errors are the authors'. A GitHub repository containing our data and code is located \href{https://github.com/NewYorkCityCouncil/cure_analysis}{here}.}}
\author {
        Rachel Avram
        \and
        Eric J. Koepcke
        \and
        Alaa Moussawi
        \and
        Melissa Nu\~nez
        }

\date{\parbox{\linewidth}{\centering%
  \today\endgraf\bigskip
  }}
  
\maketitle
\begin{abstract}
Cure Violence is a community violence intervention program that aims to reduce gun violence by mediating conflicts, ``treating'' high-risk individuals, and changing community norms. Using NYC shootings data from 2006-2023, we assess the efficacy of Cure Violence using both difference-in-differences and event study models. We find that, on average, Cure Violence is associated with a 14\% reduction in shootings relative to the counterfactual. This association persists in the years after treatment, neither increasing nor decreasing much over time. We exploit variation in the geography and timing of Cure implementation to argue against alternative explanations. Additionally, we find suggestive evidence of spillover effects into nearby precincts, as well as increasing returns to opening new Cure programs. Interpreted causally, our results imply that around 1,300 shootings were avoided between 2012-2023 due to Cure---generating a net social surplus of \$2.45 billion and achieving a benefit-cost ratio of 6.5:1.
\end{abstract}

\section{Introduction}

The Covid-19 pandemic introduced a significant uptick in gun violence, with gun homicides in the U.S. increasing 45\% between 2019-2021 (\cite{Gramlich23}). Thankfully, this spike in violence has been short-lived---e.g., in New York City (NYC) shootings and murders dropped 24.7\% and 11.9\%, respectively, from 2022-2023 (\cite{NYPD24}). Despite this, as well as the fact that crime rates are historically low, Americans remain deeply concerned about crime and gun violence---74\% of New Yorkers say crime is a ``very serious'' problem, and six-in-ten U.S. adults say gun violence is a ``very big'' problem (\cite{AkinnibiWahid22}; \cite{Schaeffer23}).

In response, politicians have directed considerable resources towards a multi-pronged approach to violence reduction that combines traditional law enforcement with community violence intervention (CVI) programs. State and local funding from the 2021 American Rescue Plan has been used by cities across the nation to start and expand CVI programs, and \$5 billion in funding for CVI programs could be on the way via the Break the Cycle of Violence Act (\cite{MacGillis23}; \cite{OMB22}). Locally, NYC's network of gun violence prevention programs now has a budget 2.5 times its FY2020 size.

Given the resources being directed towards CVI, it is critical to assess the efficacy of these programs. This paper studies Cure Violence, one of the most prominent CVI programs.\footnote{Throughout this paper, we use ``Cure Violence'' and ``Cure'' interchangeably.} Using a public health approach, Cure seeks to reduce gun violence by employing ``credible messengers'' that mediate conflicts, ``treat'' individuals at high-risk of gun violence/victimization, and change community norms. An expanding literature has analyzed Cure programs in cities throughout the U.S. and abroad. Utilizing a range of methodologies, the literature has generally found that Cure programs result in changes in gun violence that range from mixed to very significant reductions.

However, past research into Cure has been limited by small sample sizes---either in terms of the number of Cure programs, the length of time covered, or both. In contrast, our study provides the first citywide analysis of Cure's impact on gun violence. To do so, we use NYC shootings data from 2006-2023, with 28 Cure and 48 control police precincts, to estimate difference-in-differences (DID) and event study Poisson regression models with robust standard errors. Our data allow us to make a stronger causal argument and obtain more reliable estimates of Cure's impact than previous studies have been able to.

On average, we find that the presence of Cure Violence in a precinct is associated with a significant 14\% reduction in shootings relative to the counterfactual of no Cure program. We find that this reduction in shootings occurs immediately in the year of Cure implementation and persists, neither increasing nor decreasing much with time. We support our analysis by providing evidence for the ``parallel trends assumption,'' the identifying assumption behind our research design. We exploit variation in the geography and timing of Cure implementation to argue against alternative explanations for our results. Additionally, we use crime and New York Police Department (NYPD) data to argue that our findings are not due to crime trends within precincts nor due to the NYPD.

Given that Cure participants and workers are free to move around the city, past research has commented on the potential for spillover effects (e.g., \cite{SkoganEtAl09}; \cite{WebsterEtAl12}). Using a nonparametric definition of potential ``spillover treatment,'' we find evidence that neighboring precincts also experience reductions in shootings---even those that never received a Cure program themselves. Spillover effects imply a violation of the Stable Unit Treatment Value Assumption (SUTVA), however additional results suggest that this does not meaningfully bias our main findings---if anything, it attenuates them somewhat. We also find no evidence that spillover effects led to diminishing returns to opening new Cure programs---rather, the evidence suggests increasing returns to program expansion.

Past research has not conducted a cost-benefit analysis to determine if Cure Violence is a cost effective approach to reducing gun violence. This omission has been lamented by policymakers and has hindered the expansion of Cure Violence programs (\cite{ButtsEtAl15a}). Using our estimate of Cure's impact on shootings and the \textcite{LudwigCook01} estimate of the societal cost of criminal gun violence, we derive a benefit-cost ratio for Cure in NYC. Our estimate implies that, on average, NYC would have experienced 7.6\% more shootings between 2012-2023 in the absence of Cure Violence. Over those years, this adds up to around 1,300 shootings avoided. We estimate that this generated a net social surplus of \$2.45 billion, achieving a benefit-cost ratio of 6.5:1.

The remainder of the paper is laid out as follows. Section 2 provides background on Cure Violence, generally and in NYC specifically. Section 2 also briefly summarizes past research on Cure and describes how our work contributes to that literature. Section 3 provides details on our data and methodology. Section 4 provides support for the parallel trends assumption, presents estimates of Cure's impact on gun violence in NYC, and argues that our results are unlikely to be caused by something other than Cure. Section 4 also provides suggestive evidence for spillover effects and presents results that explore the statistical and policy implications of such effects. Section 5 uses our results to perform a cost-benefit analysis of Cure. Section 6 discusses limitations to our work, policy ideas, and concludes.

\section{Cure Violence}
\label{sec:cure}

\subsection*{Cure Violence Background}

Cure Violence is considered one of the most prominent CVI strategies both in the U.S. and abroad.\footnote{There are, of course, other CVI approaches besides Cure Violence---e.g., Boston CeaseFire (see \cite{PicardFritscheCerniglia13}), Chicago READI (see \cite{BhattEtAl24}), or Advance Peace (see \cite{PuglieseEtAl22}). What makes Cure unique is its community oriented, public health and prevention approach.} Originally called CeaseFire, Cure was founded by Dr. Gary Slutkin at the University of Illinois Chicago and was first launched in 1999 in Chicago's West Garfield Park. Over the years, with financial support from governments and private and community foundations, Cure Violence programs expanded to other U.S. cities. In 2008, Cure expanded internationally to Iraq and eventually spread to other cities around the globe.\footnote{For further information on the history of Cure Violence and how the program operates, see, e.g., \textcite{ButtsEtAl15a}, \textcite{CVG21}, and \textcite{SkoganEtAl09}.}

Using an epidemiological lens, Cure Violence views gun violence as a learned, transmissible behavior that can be treated and interrupted. Based on the World Health Organization's approach to stopping the spread of infectious disease, Cure Violence attempts to detect and interrupt conflicts before they escalate, identify and ``treat'' high-risk individuals (i.e., those most likely to commit or be victims of gun violence), and change community social norms surrounding gun violence. To accomplish these goals, Cure programs employ ``outreach workers'' and ``violence interrupters.'' These workers are typically individuals who grew up in or live in the Cure program's catchment area, and many of them have previously been incarcerated or belonged to a street gang. Because of their experiences, Cure workers have a higher likelihood of being seen as ``credible messengers'' by high-risk individuals. To maintain the trust of high-risk individuals, Cure Violence staff perform their work separately and independently from law enforcement, and the program's approach does not involve the use of force nor the threat of punishment.

Using a common approach to social service delivery, Cure Violence utilizes a decentralized, ``local host'' model to run Cure programs. For example, in NYC, Cure programs are overseen by the Department of Youth and Community Development (DYCD).\footnote{NYC Cure programs were formerly overseen by the Mayor's Office of Criminal Justice.} DYCD and the City government identify areas in the city with high rates of gun violence and then contract a qualified community-based organization (CBO) in the neighborhood to manage a Cure program. DYCD and the Cure Violence Global organization will provide the city's Cure programs with support and training, but the day-to-day operations are overseen by the ``local host.'' Employing CBOs ensures that Cure programs are led by individuals familiar to the community, who understand its needs and can effectively collaborate with local leaders and organizations. 

For a given Cure program, workers canvas a ``catchment area''---chosen to cover an area of clustered violence---and leverage their existing social networks and community knowledge to stay aware of potential conflicts and attempt to de-escalate conflicts before they lead to gun violence.\footnote{In one study, Cure workers estimated that the average conflict mediated involved 12 people and half of these conflicts were ``very likely'' to have otherwise ended in gun violence. Workers judged that 60\% of mediated conflicts were ``completely resolved'' and 30\% were ``temporarily resolved'' (\cite{PicardFritscheCerniglia13}).} Workers develop relationships with high-risk individuals and work with them intensively in order to dissuade them from using violence and to connect them with any social services they need (e.g., employment services, drug treatment, or housing).\footnote{E.g., \textcite{PicardFritscheCerniglia13} note that high-risk individuals fit a majority of the following criteria: (1) 16-25 years old, (2) recently released from prison, (3) recent victim of a shooting, (4) major player in a street gang/organization, (5) active in a violent street gang/organization, (6) history of violence/crime against persons, (7) carry a weapon. In their study, a majority of Cure participants were young men of color, who were gang-involved, unemployed, and had not completed high school/GED. They note that an outreach worker typically has a caseload of 5-15 high-risk individuals and spends about 20 hours one-on-one with each individual. Cure workers try to boost high-risk individuals' perceived costs of gun violence (e.g., by emphasizing risk of injury/death or incarceration and risks to their family/loved ones) and promote alternative ways to resolve conflicts that often lead to gun violence.} Staff also meet with community leaders, organizations, and residents in order to build support for and convey a message that the community will not tolerate gun violence.\footnote{Cure workers use various means to convey this message, including media campaigns, signs/billboards, and public events, such as marches, rallies, and prayer vigils after shootings.} Prior to beginning their work, staff are extensively trained as community health specialists and learn evidence-based techniques to mediate conflict, persuade, and change behaviors and norms. Cure workers also keep in-house (de-identified) data on all of the work that they do, in order to track their performance. 

\subsection*{Cure Violence in NYC}

In 2011, City Council Speaker Christine Quinn created an anti-gun violence task force with the goal of addressing gun violence through community-driven strategies. The task force identified Cure Violence as a promising, evidence-based program. In 2012, the Cure approach was first adopted in the East New York and Crown Heights neighborhoods of Brooklyn, joining existing anti-gun violence initiatives already in place there. The following year, Cure programs were opened in the Bronx and Queens. Today, Cure Violence programs have been implemented in 28 police precincts throughout all five NYC boroughs.

In 2014, Mayor Bill de Blasio and Council Speaker Melissa Mark-Viverito unveiled the Crisis Management System (CMS), a network of CBOs providing programs designed to address community needs. While Cure Violence is the primary service offered by CMS, there are also wrap-around services (e.g., legal aid, therapeutic and mental health services, and employment services) which recognize the need for a holistic approach to combating gun violence. In 2017, Mayor de Blasio established the Office to Prevent Gun Violence to provide further support to the CMS network.

We note that our analysis cannot disentangle the effects of Cure from other CMS services. Therefore, the results presented in this paper could be thought of as evidence for the impact of CMS. We frame our results as evidence for the impact of Cure and use that terminology throughout. We chose this framing for multiple reasons. First, Cure Violence is the central component of CMS, and CMS was created around the Cure Violence approach.\footnote{This is made clear by the way the City itself describes CMS---see, e.g., the NYC Office to Prevent Gun Violence \href{https://www.nyc.gov/site/peacenyc/interventions/crisis-management.page}{website}.} Second, across cities, Cure workers are trained to refer high-risk individuals to appropriate social services---thus, even in the absence of CMS, social service referrals would have been given.\footnote{E.g., see \textcite{CVG21} or \textcite{SkoganEtAl09}.} Third, if the mechanism through which Cure affects gun violence operates via referrals to social services, we would still consider this a ``Cure effect.'' Fourth, for the few Cure programs implemented \emph{before} CMS was formed, we still find that Cure is associated with significant reductions in shootings. With this said, it is up to the reader to determine for themselves how best to interpret our evidence.   
 
\subsection*{Previous Research on Cure Violence}

An expanding literature has analyzed Cure Violence programs in cities throughout the U.S. and abroad.\footnote{For NYC, see \textcite{ButtsEtAl15b}, \textcite{DelgadoEtAl17}, and \textcite{PicardFritscheCerniglia13}. For other U.S. cities, see \textcite{BuggsEtAl21}, \textcite{FoxEtAl14}, \textcite{HenryEtAl14}, \textcite{RomanEtAl17}, \textcite{SkoganEtAl09}, \textcite{WebsterEtAl12}, and \textcite{WebsterEtAl18}. For international studies, see \textcite{MaguireEtAl18} and \textcite{RansfordEtAl16}.} These studies have utilized methodologies ranging from pre-post (time series) analyses with matched comparison units, DID designs, and synthetic control designs. Via methodology and explicit controls, these studies have tried to rule out confounding explanations for changes in gun violence, such as macro trends or law enforcement activity. In general, the literature has produced suggestive evidence that Cure programs result in changes in gun violence that range from mixed to very significant reductions. Furthermore, the strongest studies---in terms of data and methodology---tend to be the ones with results suggesting significant reductions in violence. For literature reviews see \textcite{ButtsEtAl15a}, \textcite{CVG21}, and \textcite{PuglieseEtAl22}. 

Despite this promising body of research, prior analyses of Cure Violence have been limited by small sample sizes---either in terms of the number of Cure programs or the length of time covered or both.\footnote{For example, \textcite{ButtsEtAl15b} used 2010-2013 data with three treated areas, \textcite{DelgadoEtAl17} used 2009-2016 (shootings) data with two treated precincts, \textcite{FoxEtAl14} used 2007-2011 data with one treated neighborhood, \textcite{HenryEtAl14} used 2010-2013 data with two treated districts, \textcite{PicardFritscheCerniglia13} used 2010-2012 data and one treated precinct, and \textcite{WebsterEtAl12} used 2003-2010 data with four treated police posts. \textcite{BuggsEtAl21}, \textcite{RomanEtAl17}, \textcite{SkoganEtAl09}, and \textcite{WebsterEtAl18} used time series longer than a decade, but still had few treated areas.} In contrast, our study provides the first citywide analysis of Cure's impact on gun violence---using NYC shootings data from 2006-2023, with 28 treated police precincts and 48 control precincts over that time period. Our data allow us to make a stronger causal argument and obtain more reliable estimates of Cure's impact than previous studies have been able to.

Given that each Cure program is run by a local host, there will be some amount of operational variation between programs. For example, in NYC, \textcite{PuglieseEtAl22} note that while the core operations of each Cure program follow the Cure Violence principles, program designs vary somewhat across sites. Past studies have documented some implementation challenges---e.g., in \textcite{WebsterEtAl12}, two of the Baltimore neighborhoods intended for treatment lacked strong CBOs to serve as local hosts. Difficulties can also arise after a Cure program has started---e.g., \textcite{MacGillis23} notes that some Cure workers have fallen back into lives of crime, and that Cure programs in Baltimore and Louisville have been suspended, defunded, or closed because of this. Operational differences across programs could impact Cure's efficacy. In general, the literature suggests that fidelity to the Cure Violence approach is critical to successfully reducing gun violence---e.g., in \textcite{HenryEtAl14}, the Cure program that took a more ``novel'' approach appeared to be less effective than the program that took a more traditional approach.\footnote{See also \textcite{WilsonEtAl11}, which analyzed a program in Pittsburgh that was partially modeled on Cure Violence, but did not include staff conducting violence interruption.} 

Operational differences across programs are important because they imply that there is variation in the ``dosage'' of the Cure Violence treatment across areas.\footnote{Additionally, there can be natural variation \emph{within} a Cure program \emph{over time}, e.g., due to the waxing and waning of funding available.} This is a feature to be expected from Cure's local host model, not a bug. Therefore, from a policy perspective, it does not make sense to analyze Cure programs individually. Instead, the policy-relevant parameter is the \emph{average} Cure effect across all programs---which takes for granted that there will be variation in how ``well-run'' each program is. Studies with few treated areas lack the statistical power to reliably estimate an average treatment effect and short time series prevent estimation of long-run effects.\footnote{Given that Cure is meant to impact communities---not just individuals---treatment effects should be estimated at some geographic level, which also necessarily limits sample size and statistical power (\cite{ButtsEtAl15a}).} With 28 treated precincts and a long time series, we can estimate average Cure treatment effects in the short- and long-run more reliably and with better precision than past studies.

Aside from improved statistical power, our data allow us to make a strong causal argument for Cure. As mentioned, while the literature has continually found at least some evidence that Cure reduces gun violence, limited data has prevented previous studies from convincingly ruling out confounding explanations, such as macro or regional crime trends. For example, in \textcite{WebsterEtAl12}, three of the four treated Baltimore neighborhoods are clustered together, meaning that some localized change in violence could explain any association between Cure and gun violence in that area.\footnote{Indeed, \textcite{WebsterEtAl12} note that an intense gang feud occurred in that area during the time of Cure treatment and likely led to a spike in homicides.} Additionally, all four treated neighborhoods in \textcite{WebsterEtAl12} received Cure in 2007 or 2008, meaning that time trends in gun violence could confound their results.\footnote{For another similar example, see \textcite{RomanEtAl17}.} In contrast, the treated precincts in our data are spread across NYC (in all five boroughs) and have treatment dates that range from 2012-2021. While not providing experimental evidence for Cure's impact, this geographic and temporal variation in treatment significantly reduces the chance that our results can be explained by alternate explanations. Additionally, we use felony crime and NYPD headcount, budget, and felony arrest data to argue that our findings are not due to crime trends within precincts nor due to the NYPD. 

Previous research has suggested that the effects of Cure can \emph{spill over} into surrounding areas. For example, \textcite{SkoganEtAl09} found that only half of the Cure participants they surveyed lived in or frequented a Cure catchment area and that program paperwork showed that Cure staff conducted work over broad areas. Only one previous study has actually attempted to estimate these spillover effects---\textcite{WebsterEtAl12} found reductions in gun violence in bordering areas that were of similar magnitude to the reductions seen in  treated neighborhoods. We estimate spillover effects in a similar manner, contributing more precise and robust evidence for spillover effects than has previously been seen.

Spillovers raise important statistical and policy implications. If one ignores spillover effects, then they may underestimate Cure's impact by not counting Cure's effect on nearby areas and by using control units that receive some treatment themselves. Studies using a matching or synthetic control design must be careful, as areas adjacent to treated neighborhoods are likely to be similar to treated areas and could be spilled into.\footnote{E.g., in \textcite{SkoganEtAl09} and \textcite{PicardFritscheCerniglia13}, most or all of the control units used are contiguous to treated units.} If Cure spills over into surrounding areas, then there may be diminishing returns to expanding the number of Cure programs. Or, perhaps, collaboration between nearby Cure sites may lead to increasing returns to scale (\cite{ButtsEtAl15a}). Previous research has not formally investigated whether spillovers are biasing estimates of Cure's impact nor what the returns to expanding Cure sites are. Our data allow us to analyze these questions and contribute the first suggestive answers to them.

\section{Data and Methodology}
\label{sec:data}
\subsection{Data}
\subsubsection*{Cure Violence Programs}

Cure Violence programs have been implemented and are operational across all five NYC boroughs. The Mayor's Office of Criminal Justice (MOCJ) gave us information on each NYC Cure program as of January 2022.

As noted, Cure programs technically operate in a small catchment area. However, Cure operations and people can spill over catchment borders. Additionally, if gun violence activity moves, catchment area borders can shift somewhat over time. Given this and the fact that Cure is meant to impact entire communities, we use police precincts as our geographic units. We note that police precincts are much larger than Cure catchment areas and, thus, we may be over-attributing treatment. If so, this would lead us to underestimate Cure's impact on gun violence.

MOCJ provided us with information on 35 Cure programs. There are six police precincts that have or had multiple Cure programs.\footnote{In five of these cases, the precinct has multiple programs running concurrently (or were scheduled to start later in 2022) serving different areas of the precinct. In the other case, an earlier Cure program had shut down and a new one has replaced it.} In each of these instances, we use the earliest start date as our Cure ``treatment'' date for that precinct. This choice helps to avoid concerns about precincts receiving treatment earlier than their ``treatment date.''

\begin{figure}[!htbp]
    \centering
    \caption{Average Shootings Across NYC Police Precincts}\label{fig:avgShootingsMap}
    \includegraphics[width=\textwidth]{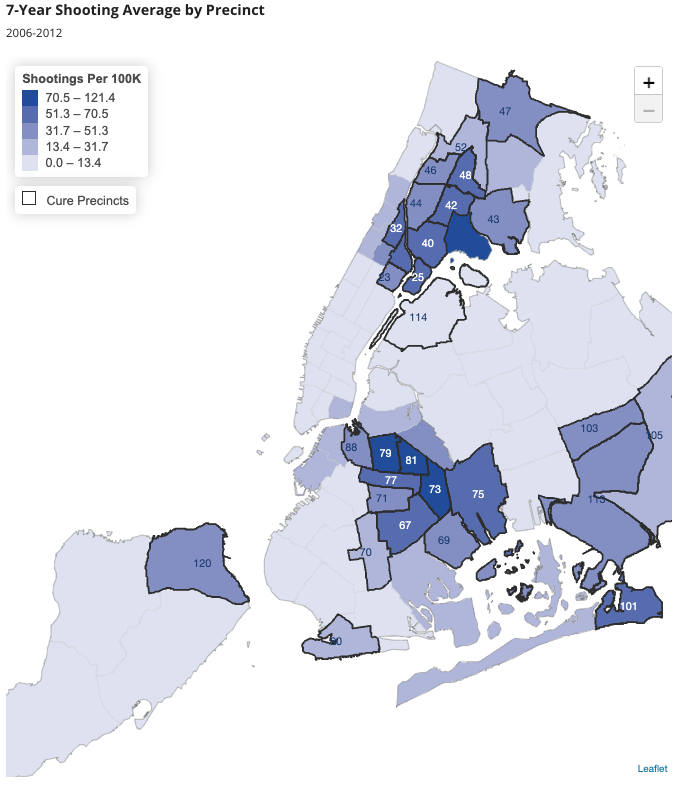}
    \raggedright
    \small{\textit{Note}: This figure provides a police precinct map of NYC. As can be seen in the legend, the color gradient signifies the seven-year (2006-2012) average shooting rate per 100,000 people in a precinct. Cure treated precincts are outlined in black.}
\end{figure}

\begin{figure}[!htbp]
    \centering
    \caption{Cure Treatment Dates Histogram}\label{fig:cureDatesHist}
    \includegraphics[width=\textwidth]{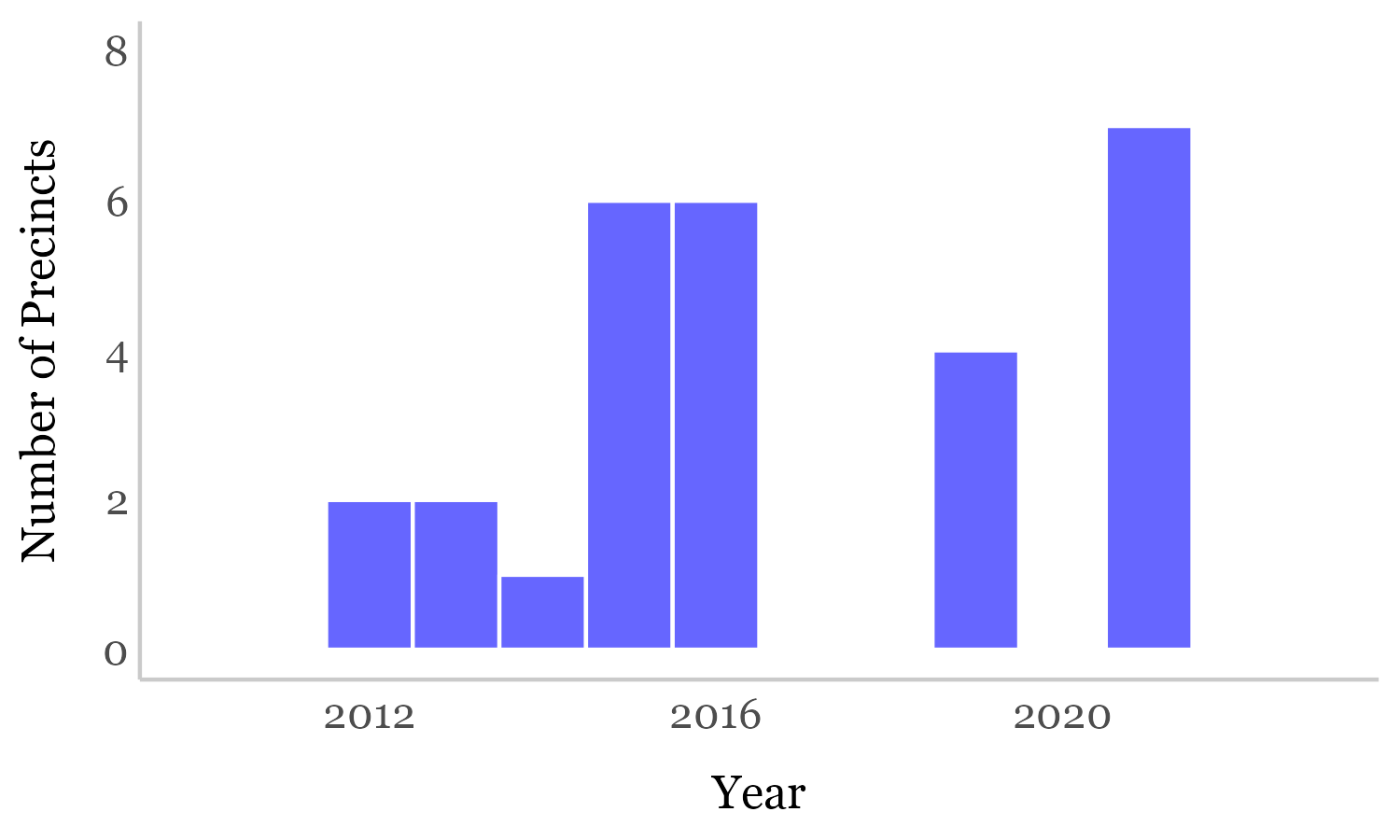}
    \raggedright
    \small{\textit{Note}: This figure plots a histogram of the Cure treatment years for the 28 Cure treated precincts. For precincts with multiple Cure programs, the earliest treatment date is used.}
\end{figure}

In total, 28 police precincts have at least one Cure program. There was a staggered rollout of Cure programs across precincts, with treatment dates ranging from 2012 to 2021. Figure \ref{fig:avgShootingsMap} provides a map of where the treated precincts are located, and Figure \ref{fig:cureDatesHist} provides a histogram of the Cure implementation dates by year. Of these precincts, two have Cure programs that started in 2021 and were not fully operational as of January 2022. Another two precincts had Cure programs that were no longer operational as of January 2022 (one of which was getting a new program later in 2022). For simplicity, we treat \emph{all} Cure programs as fully operational in our analysis, again likely leading to underestimation of any treatment effect.

Despite inevitable operational differences across Cure programs, we assume that the Cure treatment is homogeneous across precincts and we estimate an average treatment effect---averaging over how ``well-run'' each program is (and other unmodeled variation). 

\subsubsection*{Gun Violence Data}

Our outcome of interest is the number of shootings within a precinct. To measure this, we utilize the NYPD's Shooting Incident dataset.\footnote{The data is public and is available \href{https://data.cityofnewyork.us/Public-Safety/NYPD-Shooting-Incident-Data-Historic-/833y-fsy8}{here} on NYC Open Data.} The data contains information on every shooting incident that resulted in an injured victim from 2006 to the end of 2023. As with any data of this nature, our findings and conclusions assume that our data is representative of all the shootings in NYC or, at least, that the ``representativeness'' of the data did not change over time.

In the data, there are 28,562 shooting incidents recorded. There are some duplicate incident keys in the data, which occurs when there are multiple victims for a given shooting incident. As we are concerned with Cure's effect on the number of gun victims, we do not drop these duplicates. We drop precinct 22, i.e., Central Park, which only has one shooting recorded in the data.\footnote{This is likely due to the fact that (as per the data notes) shootings in open areas (like a park) are coded as occurring on the nearest street bordering the area.}  

\begin{figure}[!htbp]
    \centering
    \caption{Citywide Number of Shootings Over Time}\label{fig:numShootings_byYear}
    \includegraphics[width=\textwidth]{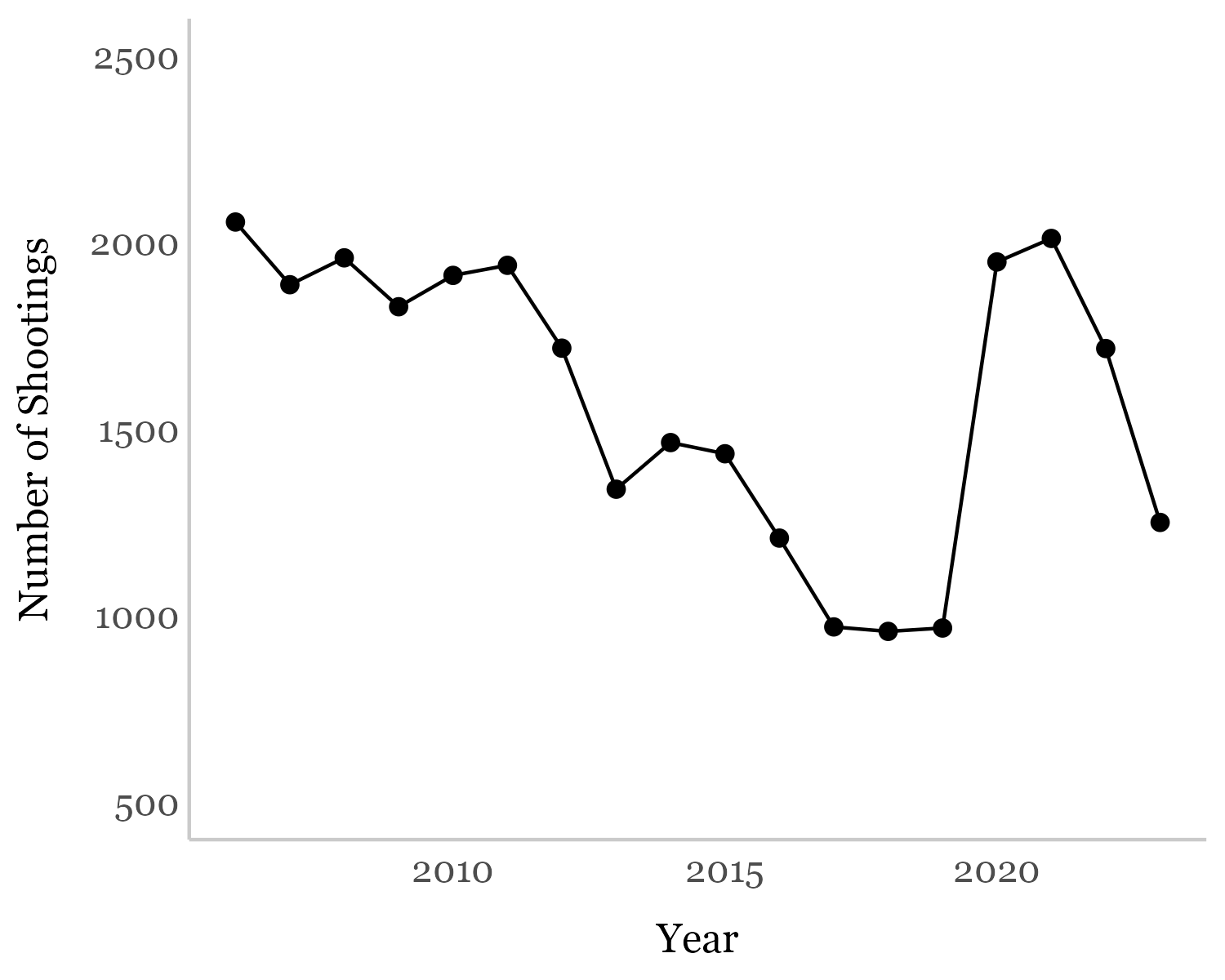}
    \raggedright
    \small{\textit{Note}: This figure plots the total number of shootings citywide from 2006-2023.}
\end{figure}

Figure \ref{fig:numShootings_byYear} plots the number of shootings citywide from 2006-2023. On average, there were around 1,600 shootings per year between 2006-2023, and the number of shootings ranges from about 1,000 to 2,000 per year. As one would expect, the number of shootings varies greatly by geography within the city (see Figure \ref{fig:avgShootingsMap}). Approximately 19\% of shootings end in a homicide. A large majority of the victims in the data are young men of color. This is true for the perpetrators as well, although there is a significant number of missing data for perpetrator descriptors.

We use police precincts as our geographic unit and calendar year as our unit of time, and we collapse the data by counting the number of shootings per precinct-year. This leaves a balanced panel of 1,368 observations for the 76 precincts across 2006-2023.

\subsection{Analytical Framework}

To assess the impact of Cure on gun violence we use a DID design. In the language of \textcite{Miller23}, our context provides us with a ``hybrid'' data structure, where we have both treated and control police precincts and a staggered rollout of Cure programs across treated precincts. Therefore, our identifying variation for treatment effects comes from these two sources: (1) the comparison of the treated and control precincts' time trends and (2) the comparison of earlier- and later-treated precincts' time trends. There are 28 Cure treated precincts and 48 control precincts (see Figure \ref{fig:avgShootingsMap} and Figure \ref{fig:cureDatesHist}).

The critical assumption of DID designs is the ``parallel trends assumption''---i.e., in the absence of Cure, the average number of shootings in control and (earlier/later) treated precincts would have followed parallel time trends. This assumption would be satisfied if the Cure precincts were randomly chosen and if the timing of Cure implementation had been randomized. However, the Cure precincts were not randomly selected---they were chosen for Cure because of their high levels of gun violence. Fortunately, randomization is not necessary---level differences between treated and control precincts can be accounted for, we just require that their time trends would have been parallel in the absence of treatment. Section 4 presents evidence for the parallel trends assumption.

Our outcome variable (number of shootings) is a non-negative integer, which is zero in 7\% of our precinct-year observations. Additionally, the level of shootings varies significantly across the city and, thus, we would like a relative measure of Cure's impact on gun violence, rather than an absolute one. Given these considerations, we use Poisson regression models in our analyses. We note that the shootings data is over-dispersed. However, as with ordinary least squares, Poisson regression is very general and robust. In particular, Poisson regression does \emph{not} require a Poisson distributed outcome variable for unbiased parameter estimates---only the conditional expectation function needs to be correctly specified (i.e., it requires exogeneity).\footnote{See \textcite{ChenRoth24}, \textcite{Gould11}, \textcite{Hansen22}, \textcite{Winkelmann15}, and \textcite{Wooldridge10}. As noted in some of those references, Negative Binomial regression is not nearly as robust as Poisson regression.} However, if the outcome variable is not Poisson distributed, the standard errors will be incorrect. To correct for this, we use robust standard errors in all analyses.  

In our framework, our units are police precincts, and our calendar time periods are years, which we index by $i$ and $t$, respectively. Our Poisson DID models take the form:
\[
log(E[y_{it}|\alpha_i, \delta_t, \mathbf{T_{it}}]) = \alpha_i + \delta_t + \mathbf{\gamma T_{it}}
\]
where $y_{it}$ is the number of shootings in precinct $i$ in year $t$ and $\alpha_i$ and $\delta_t$ are precinct and year fixed effects, respectively. 

We vary the Cure treatment vector, $\mathbf{T_{it}}$, across specifications to allow for more/less flexibility in treatment effects across time. In our simplest specification, $\mathbf{T_{it}}$ only contains one binary indicator that is ``on'' once a precinct receives Cure. In this case, our parameter of interest, $\gamma$, captures the average Cure treatment effect, not allowing for the effect to change with time. A second specification includes a variable counting the number of years a precinct has had a Cure program. In this specification, the coefficient on the binary treatment variable captures the immediate or base effect of Cure, and the coefficient on the ``years of Cure'' variable measures a linear dynamic treatment effect. A third specification additionally includes a squared ``years of Cure'' variable, which allows for a quadratic dynamic treatment effect. The latter two specifications allow us to see whether treatment effects are persistent or increase/decrease with time---a key consideration when policymakers are making decisions about funding Cure.
 
To allow for further flexibility in the Cure treatment effects across time and to more formally check the parallel trends assumption, we also use event study specifications of the form:
\[
log(E[y_{it}|\alpha_i, \delta_t, D_{it}^k]) = \alpha_i + \delta_t + \sum_{k=-6}^3 \gamma_k D_{it}^k
\]

The $D_{it}^k$s are indicator variables, which connect ``event time,'' indexed by $k$, to calendar time for a given precinct. Specifically, each treated precinct has a Cure start year, $s_i$, and $D_{it}^k = 1\{t=s_i+k\}$.\footnote{E.g., say that precinct $i$ received Cure in 2015 and an observation's calendar year is $t=2018$, then the event time is three years post-treatment and $D_{it}^3=1$ and all other $D_{it}^k$s are zero. Similarly, if the calendar year is 2015 then $D_{it}^0=1$ (and all others are 0), and if the calendar year is 2013, then $D_{it}^{-2}=1$ (and all others are 0).} Our shootings data cover 2006-2023 and our Cure start dates range from 2012-2021. We would like for each coefficient, $\gamma_k$, to be estimated off the full set of treated precincts, but we are also interested in measuring long-term effects. With these competing considerations in mind, we chose to ``pool'' event times outside of the window $k \in [-6,3]$ in end-cap indicators. Specifically, we have $D_{it}^3 = 1\{t \geq s_i+3\}$, so that all Cure precinct observations with event times equal to or greater than three years post-treatment have $D_{it}^3 = 1$.\footnote{Note that $\gamma_3$ can only be estimated using precincts treated between 2012-2019. The remaining $\gamma_k$s are estimated using all treated precincts. We judged that adding an extra year post-treatment effect was worth this tradeoff.} Similarly, we have $D_{it}^{-6} = 1\{t \leq s_i-6\}$, so that all Cure precinct observations with event times six years pre-treatment or more have $D_{it}^{-6} = 1$.\footnote{With these pooled indicators, every treated precinct observation has exactly one $D_{it}^k$ turned on (i.e., equal to one). For control precincts, all $D_{it}^k$s are zero in every year.}

The coefficients of interest are the $\gamma_k$s, which capture the dynamic ``treatment effects'' of Cure for the years leading up to and after Cure implementation. As is standard, we omit the first pre-treatment indicator (i.e., $D_{it}^{-1}$) from regressions to use as our reference period in event time.\footnote{Precinct 1 and 2006 are the omitted reference precinct and year in regressions, respectively.} Thus, the treatment coefficients have a DID interpretation---i.e., they give the average difference in shootings between treated and control precincts in event time $k$ relative to that difference in event time $-1$.\footnote{The interpretation of the coefficients on the pooled end-cap indicators is slightly different---$\gamma_{-6}$ gives the average ``treatment effect'' for Cure precincts six years or more pre-treatment and $\gamma_{3}$ gives the average treatment effect for Cure precincts three years or more post-treatment.}

The event study specification allows for complete flexibility in the dynamic treatment effects of Cure. The immediate treatment effect of Cure is captured by $\gamma_0$ and the longer term treatment effects are captured by the post-treatment coefficients, $\gamma_k$ with $k>0$.\footnote{Of course, due to the pooling of event times $k\geq 3$, our specification cannot capture nuance in dynamic treatment effects beyond this point.}

The pre-treatment $\gamma_k$s with $k<0$ serve as placebos---as Cure has not yet been implemented, we would hope to find that they are insignificant. If this is not the case, it may indicate some anticipatory effect or model misspecification. If the pre-treatment coefficients are insignificant, then this improves our confidence in the parallel trends assumption, because it indicates that treated and control precincts were following parallel trends prior to Cure implementation.

\subsubsection*{Spillover Effects}

Given the geographic proximity of police precincts, we explore the possibility of spillover effects of Cure between precincts. Past research has commented on this possibility (\cite{SkoganEtAl09}; \cite{WebsterEtAl12}) and, as we will see, there is suggestive evidence of this in our data. 

Spillover effects complicate the measurement of treatment effects, as their presence would invalidate the parallel trends assumption. Specifically, spillover effects imply that SUTVA is violated---i.e., some control precincts or later-treated Cure precincts might be receiving some amount of treatment via spillover and, thus, may no longer tell us what the time trend would have been in the absence of treatment. If this is the case, our estimates may suffer from attenuation bias, making it more difficult to identify a treatment effect. Additionally, any spillover between earlier- and later-treated precincts could complicate the use of pre-trends as a test for the parallel trends assumption and complicate the analysis of long-term treatment effects.

It should be noted that these issues are not unique to the DID methodology---spillover effects would also bias analyses utilizing pre-post or synthetic control designs.\footnote{Spillover between earlier- and later-treated precincts would bias pre-post designs. In a synthetic control design, the control precincts most similar to a treated precinct are likely to be those near the treated precinct and, thus, those most likely to experience spillover effects.} To get around this issue, one must put careful thought into which precincts are being considered control and treated units.

In Section 4, we explore the possibility of spillover effects. To do so, we assume that Cure treatment can potentially spill over into \emph{any} precinct that \emph{borders} the treated precinct, but not beyond this. Of course, if spillover extends farther than this, then our ``new'' control units would still experience some spillover and we would run into the issues described above. Instead, if spillover effects do not extend quite so far, then we will be overly restrictive on how we define control units and we might be excluding control units that are the best counterfactual for treated units. We chose our nonparametric spillover assumption to balance these two concerns and be operational given our data.

Like Cure treatment, we assume that any spillover treatment is homogeneous. This allows us to define any precinct that could potentially be spilled into as a ``spillover precinct.'' Importantly, spillover precincts can be both control and treated precincts. Additionally, for Cure precincts, we allow for the ``spillover treatment'' to be distinct from the Cure treatment. Thus, our framework allows for the possibility of spillover effects from treated-to-control precincts, from earlier-to-later treated precincts, and from later-to-earlier treated precincts. Because a spillover precinct can potentially be spilled into by multiple precincts, we define a precinct's ``spillover treatment date'' as the earliest year in which a contiguous precinct received Cure.

With this framework, we can partition the precincts by their treatments. We have a total of 76 precincts, with 48 control precincts and 28 Cure precincts. In terms of spillover treatment, 24 precincts do not border a Cure precinct (20 of these are control precincts and four are Cure precincts), and 52 are spillover precincts (28 of these are control precincts and 24 are Cure precincts).

\section{Results}
\label{sec:results}

In this section, we provide evidence regarding Cure's impact on gun violence in NYC. We begin by evaluating the evidence for the parallel trends assumption. Then, we present our estimates of Cure's impact, assuming no spillover effects, and argue against alternative explanations for our findings. Finally, we provide suggestive evidence of spillover effects and present results that aim to address and understand the implications of these effects.

\subsection*{Parallel Trends Assumption}

\begin{figure}[!htbp]
    \centering
    \caption{Number of Shootings Over Time, by Treatment}\label{fig:numShootings_byCure}
    \includegraphics[width=\textwidth]{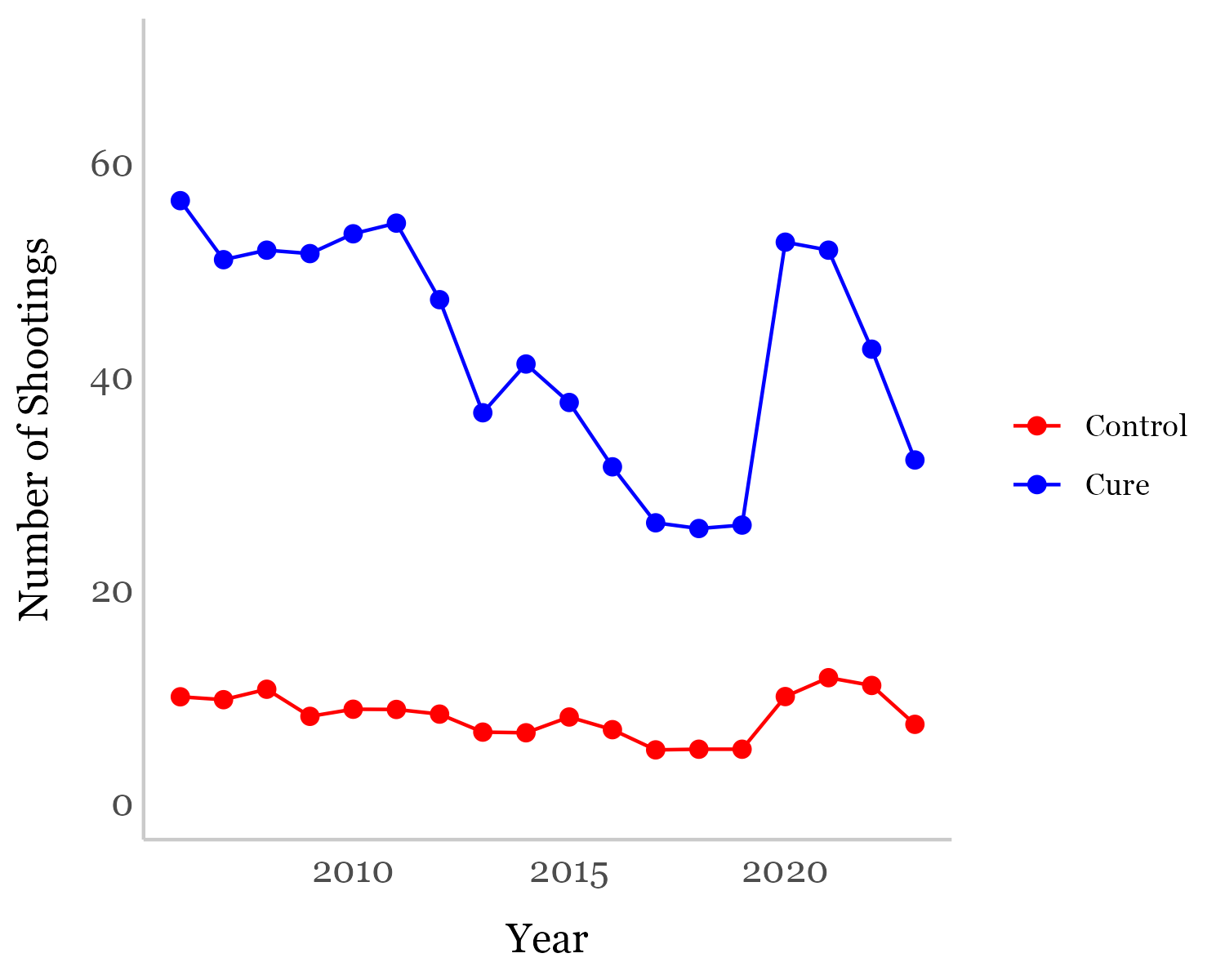}
    \raggedright
    \small{\textit{Note}: This figure plots the average number of shootings (precinct-level) over time for control precincts and Cure precincts.}
\end{figure}

Over the 2006-2023 period, there was an average of 21 shootings per precinct per year. Figure \ref{fig:numShootings_byCure} presents the time series of the average number of shootings per precinct, and shows that Cure and control precincts have an average of 43 and 8 shootings per year, respectively. Figure \ref{fig:numShootings_byCure} and Figure \ref{fig:avgShootingsMap} make clear that Cure precincts were not randomly chosen---instead, Cure precincts were chosen for treatment because of their high level of shootings. 

Table \ref{tab:balanceTable} is a balance table between Cure and control precincts using 2017-2021 American Community Survey data. We see no difference between Cure and control precincts in terms of population size. However, there are difference in racial composition---Cure precincts have a much higher percentage of Black or African American residents and a much lower percentage of white and Asian residents. Relative to control precincts, Cure precincts have lower levels of education and median household incomes. Additionally, Cure precincts have higher levels of unemployment and residents are more likely to be on public health insurance plans.

Despite the lack of randomization and differences in characteristics, Figure \ref{fig:numShootings_byCure} shows that both Cure and control precincts experienced similar trends in shootings over time---providing evidence for the parallel trends assumption. In the period between 2006-2011, before any Cure treatment, both Cure and control precincts followed a relatively flat trend line. From 2012-2019, as Cure was being rolled out, both Cure and control precincts experienced a decreasing trend in the number of shootings. Finally, the onset of Covid introduced a spike in shootings in both types of precincts, which has been rapidly tapering off as the pandemic recedes.  

\begin{figure}[!htbp]
    \centering
    \caption{Log Number of Shootings Over Time, by Treatment}\label{fig:logNumShootings_byCure}
    \includegraphics[width=\textwidth]{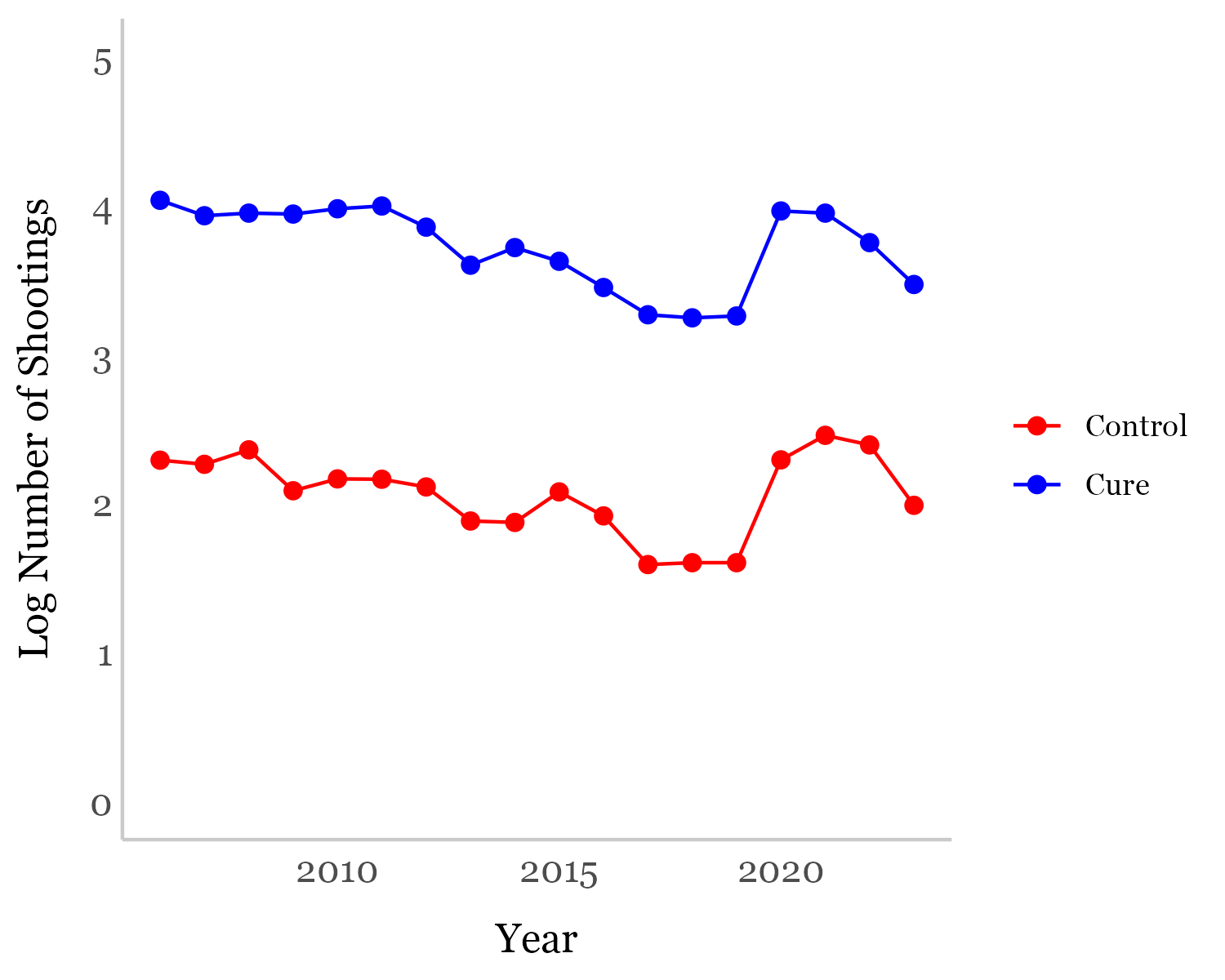}
    \raggedright
    \small{\textit{Note}: This figure plots the log average number of shootings (precinct-level) over time for control precincts and Cure precincts.}
\end{figure}

Figure \ref{fig:logNumShootings_byCure} provides additional evidence for the parallel trends assumption. The \emph{log} average number of shootings is used by our Poisson regressions, and Figure \ref{fig:logNumShootings_byCure} plots the time series of this variable for Cure and control precincts. Visually, despite a difference in levels, it is plausible that the number of shootings in Cure and control precincts would have followed parallel time trends. Our event study results below provide more formal evidence for the parallel trends assumption.

\subsection*{Impact of Cure}

Using the DID models described in Section 3, Table \ref{tab:mainPoisDIDsTab} presents regression analyses of the impact of Cure. In our simplest model, the binary Cure treatment variable is associated with a significant 14\% reduction in shootings relative to the counterfactual of no Cure treatment. The more flexible models do not show much in the way of linear or quadratic dynamic treatment effects---i.e., the results suggest that the reduction in gun violence persists in the years after treatment, neither increasing nor decreasing much with time.

\begin{figure}[!htbp]
    \centering
    \caption{Impact of Cure Event Study}\label{fig:mainPoisES}
    \includegraphics[width=\textwidth]{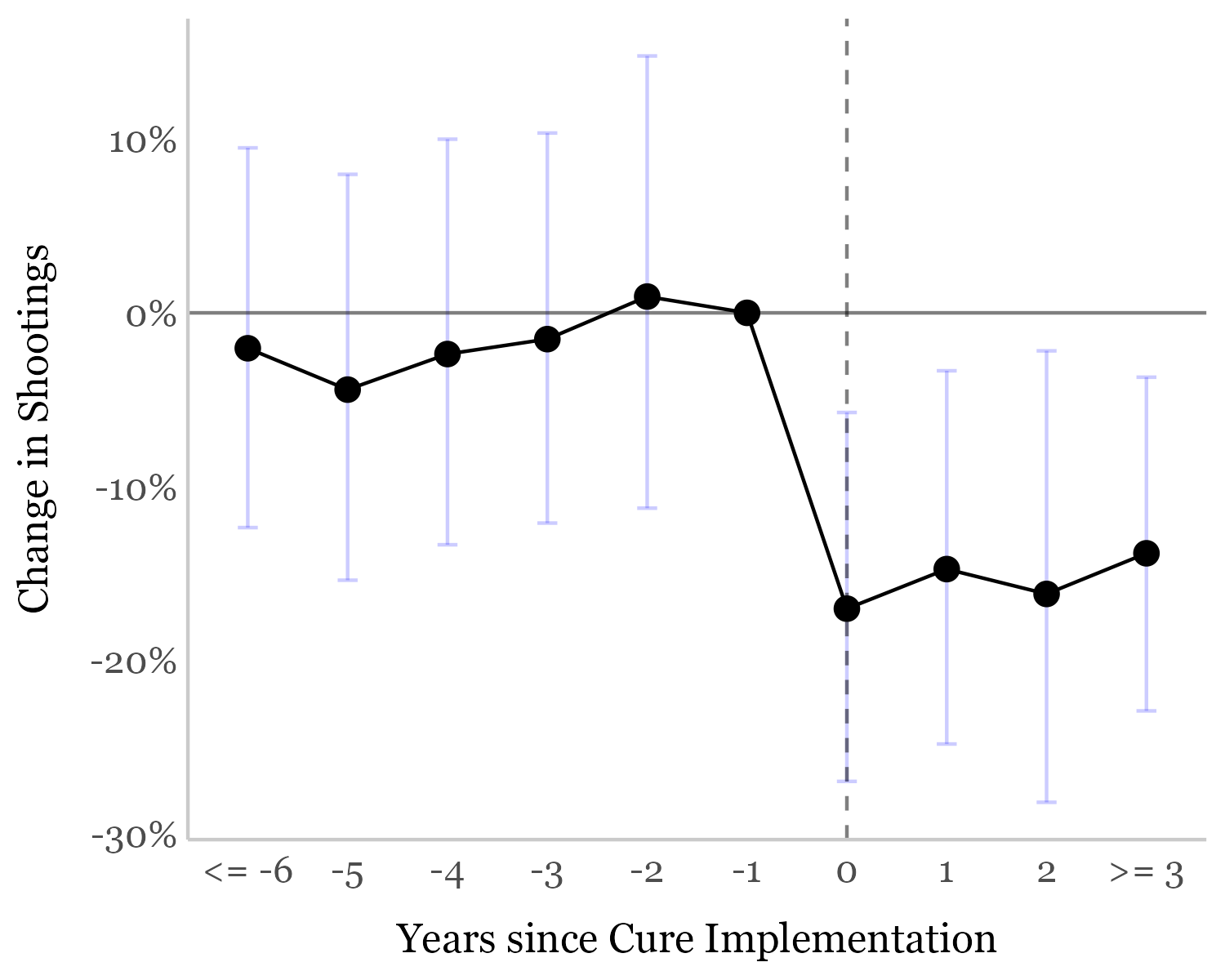}
    \raggedright
    \small{\textit{Note}: This figure plots event study treatment coefficients, using the setup described in Section 3. The first year pre-Cure is the reference period. 95\% confidence intervals are shown.}
\end{figure}

Figure \ref{fig:mainPoisES} plots the treatment coefficients from our event study model.\footnote{Table \ref{tab:mainPoisESTab} provides the corresponding regression table for Figure \ref{fig:mainPoisES}.} In the year of Cure implementation, the treated precincts experienced a significant 17\% reduction in the gap in shootings between Cure and control precincts. Confirming the results above, this reduction in shootings does not dissipate with time---the post-treatment coefficients suggest a continued significant 14-16\% reduction in shootings, even in the end-cap coefficient for three or more years post-treatment. We note that the magnitude of our estimated treatment effects are similar to those estimated by Skogen et al. (2009) in Chicago and that the ``instant and persistent'' reduction in gun violence after Cure treatment has been documented in previous studies, as well (\cite{HenryEtAl14}; \cite{SkoganEtAl09}; \cite{WebsterEtAl12}).

The pre-treatment coefficients in Figure \ref{fig:mainPoisES} are reassuring. None of these coefficients are significantly different from zero, providing more formal evidence that shootings in Cure and control precincts were following roughly parallel time trends prior to treatment, as we saw visually in Figure \ref{fig:numShootings_byCure}. These near-zero pre-treatment coefficients make the drop in shootings at the exact time of Cure treatment all the more stark. While the data does not come from a randomized experiment, the evidence \emph{strongly suggests} that Cure causes an immediate reduction in gun violence and that this persists in the years after treatment.

\subsection*{Alternate Explanations}

Next, we argue against alternative explanations for the reduction in gun violence we see in our results.

\begin{figure}[!htbp]
    \centering
    \caption{Impact of Cure: Only Pre-Covid Data}\label{fig:preCovidPoisES}
    \includegraphics[width=\textwidth]{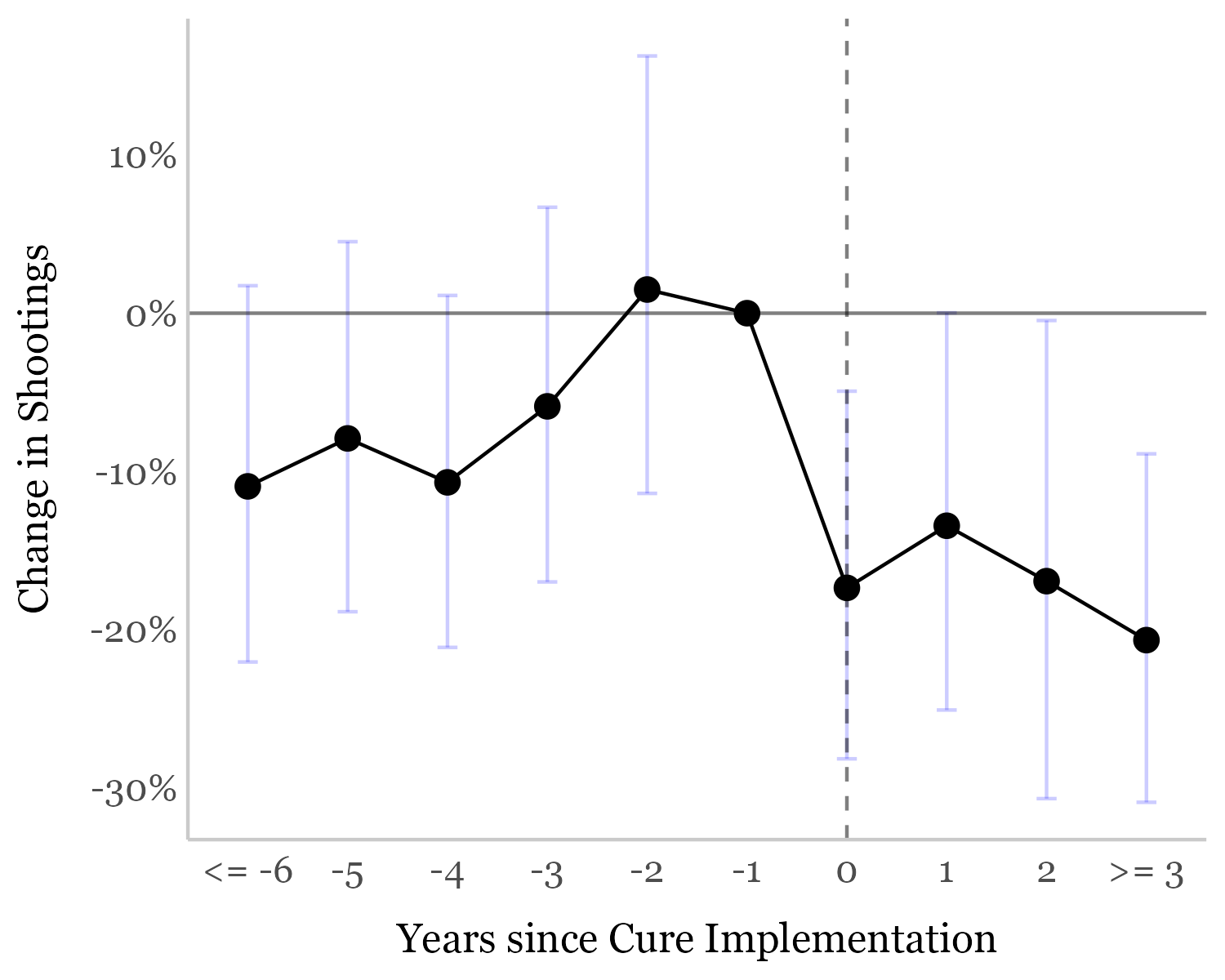}
    \raggedright
    \small{\textit{Note}: This figure plots event study treatment coefficients, using the setup described in Section 3. The first year pre-Cure is the reference period. This figure only uses shootings data from 2006-2019 and excludes Cure precincts treated between 2019-2021. 95\% confidence intervals are shown.}
\end{figure}

First, the Covid pandemic was associated with a large increase in gun violence, a fact which can be seen in Figure \ref{fig:numShootings_byCure} and Figure \ref{fig:numShootings_byYear}. Covid may have impacted how Cure programs operated, thus influencing our findings in some way. Table \ref{tab:preCovidPoisDIDsTab} and Figure \ref{fig:preCovidPoisES} present analogous DID and event study results, respectively, that only use shootings data from 2006-2019 and exclude Cure precincts treated between 2019-2021. The results are similar to those from the full dataset, providing reassurance that post-Covid data is not driving our findings.

Second, we consider the possibility of a citywide or macro shock. Such a shock would lead to similar time trends between Cure and control precincts. If this was the case, because we are accounting for the time trend in our regressions, we would not expect to see Cure treatment effects in our results. Table \ref{tab:mainPoisDIDsTab} and Figure \ref{fig:mainPoisES} clearly show significant treatment effects, ruling out this possibility. While the city did experience a downtrend in gun violence before 2020, the reduction was sharper in Cure precincts. Furthermore, the sharp reduction in shootings in Cure precincts began in 2012, the exact time when Cure began rolling out to precincts, lending further credence to the claim that this reduction was due to Cure.

\begin{figure}[!htbp]
    \centering
    \caption{Number of Shootings Over Time, by Cure Treatment Year}\label{fig:numShootings_byYearCure}
    \includegraphics[width=\textwidth]{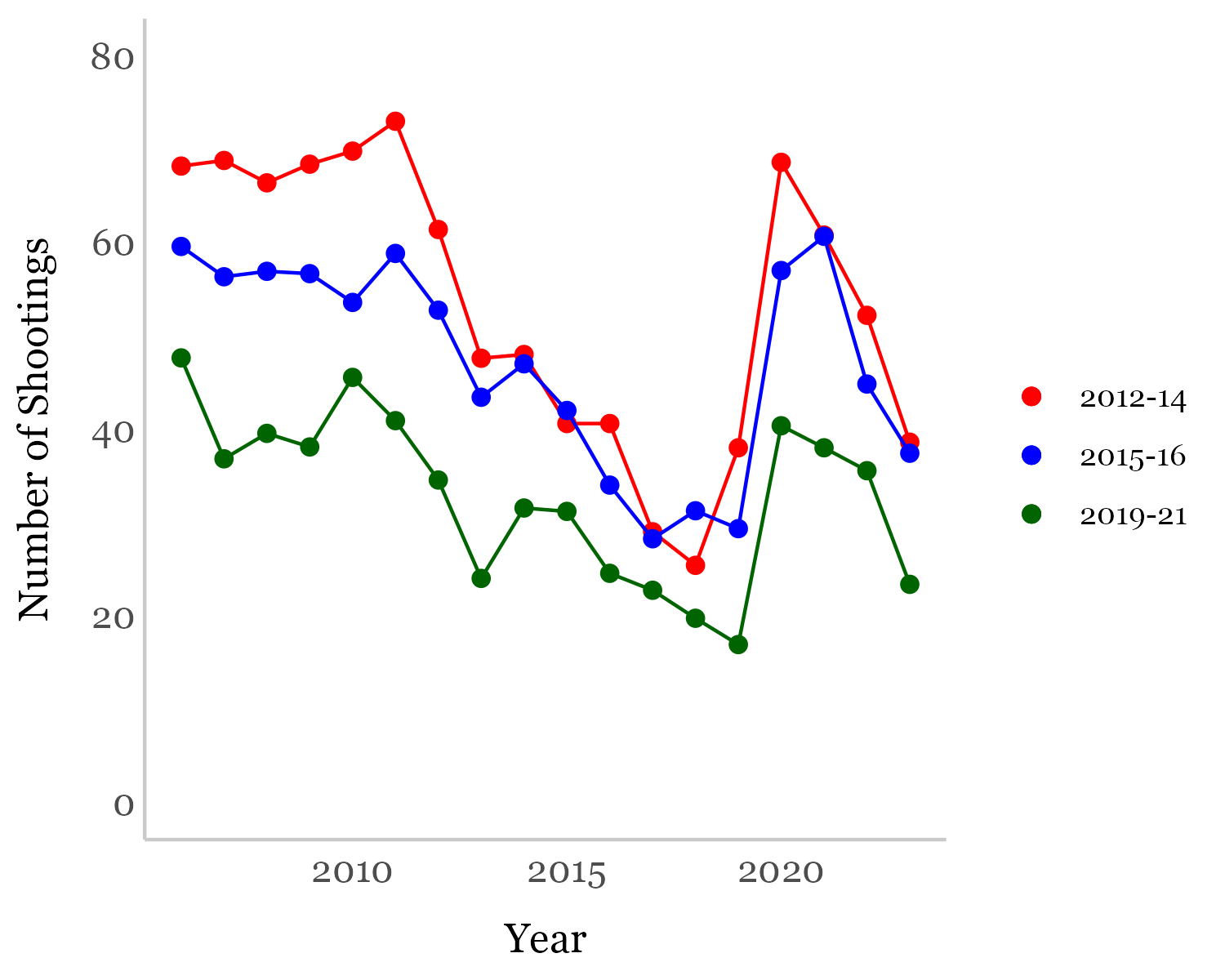}
    \raggedright
    \small{\textit{Note}: This figure plots the average number of shootings (precinct-level) over time grouped by Cure precincts treated between 2012-14, between 2015-16, and between 2019-21.}
\end{figure}

\begin{figure}[!htbp]
    \centering
    \caption{Log Number of Shootings Over Time, by Cure Treatment Year}\label{fig:logNumShootings_byYearCure}
    \includegraphics[width=\textwidth]{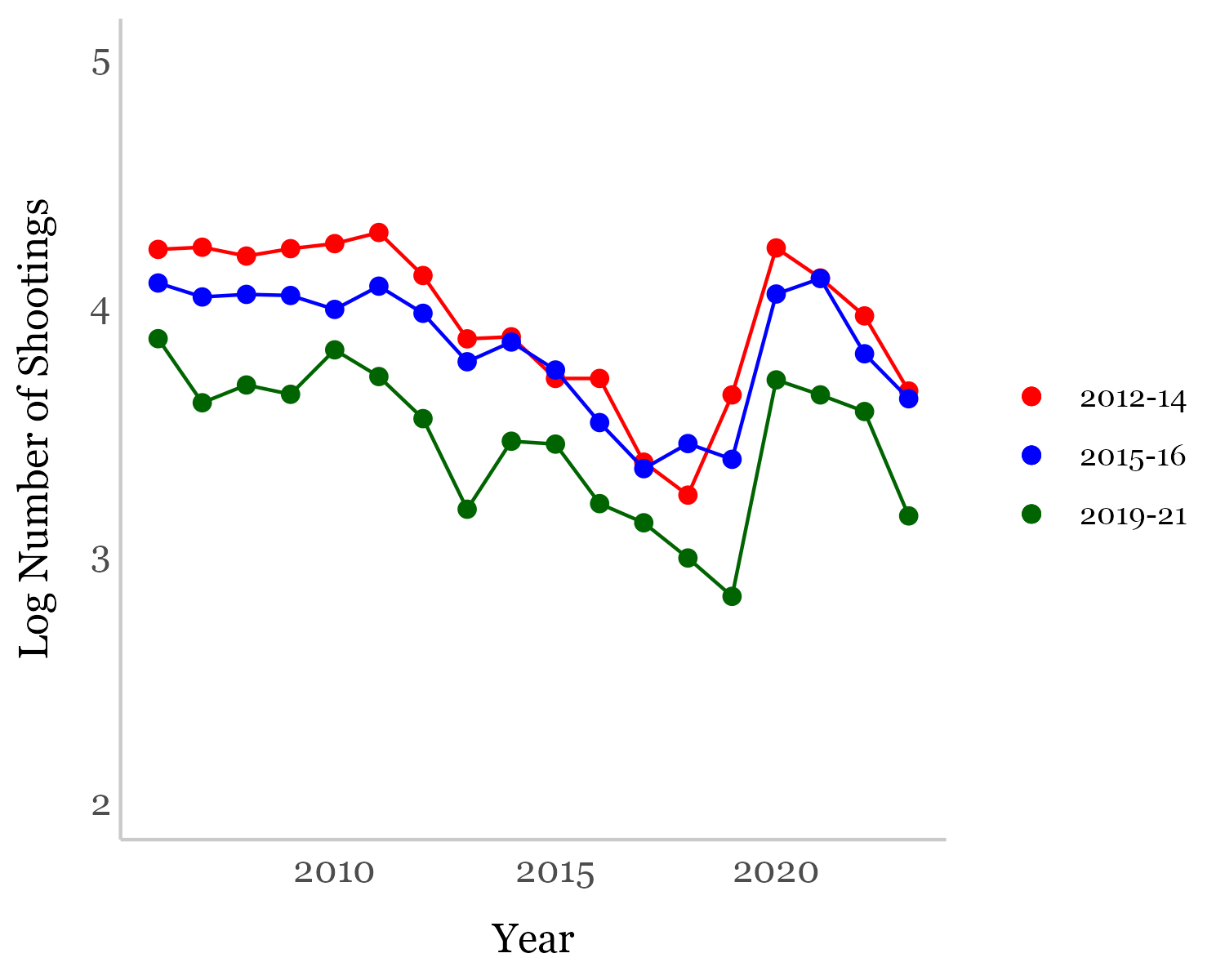}
    \raggedright
    \small{\textit{Note}: This figure plots the log average number of shootings (precinct-level) over time grouped by Cure precincts treated between 2012-14, between 2015-16, and between 2019-21.}
\end{figure}

Third, we consider a shock local to Cure precincts that happened to occur around 2012. If this was the case, then we would expect all Cure precincts to follow similar time trends, and we would not expect to find any treatment effect if we ran a regression with data containing \emph{only} Cure precincts. When dropping control precincts from the sample, treatment effects are estimated by comparing earlier- and later-treated Cure precincts under the parallel trends assumption. Figure \ref{fig:numShootings_byYearCure} plots the time series of the average number of shootings per precinct by Cure treatment date. The figure makes clear that precincts with higher gun violence were chosen for earlier treatment, however the parallel trends assumption appears plausible.\footnote{The parallel trends assumption also looks plausible in Figure \ref{fig:logNumShootings_byYearCure}, which is analogous to Figure \ref{fig:numShootings_byYearCure} but plots the times series of the \emph{log} average number of shootings per precinct.} Table \ref{tab:onlyCurePoisDIDsTab} presents DID results using only Cure precincts. Despite the fact that the Cure-only sample is around one-third the size of the full sample, Table \ref{tab:onlyCurePoisDIDsTab} still suggests that Cure reduces shootings. Our binary Cure treatment variable is associated with a marginally significant 6.5\% reduction in shootings. There also appear to be meaningful dynamic effects---our quadratic specification suggests a significant 10.5\% base reduction in shootings that wanes over time. In the next section, we provide evidence for spillover effects, which imply that our estimates in Table \ref{tab:onlyCurePoisDIDsTab} may be attenuated. Regardless, Table \ref{tab:onlyCurePoisDIDsTab} provides reassurance that our findings are not just the result of a common time trend within Cure precincts.

Fourth, we consider shocks local to Cure precincts that coincide with Cure treatment within a precinct. We note that the temporal variation in treatment makes it unlikely that such shocks could explain our results. Table \ref{tab:earlyCurePoisDIDsTab}, Table \ref{tab:midCurePoisDIDsTab}, and Table \ref{tab:lateCurePoisDIDsTab} provide DID results using Cure precincts only treated between 2012-2014, 2015-2016, and 2019-2021, respectively.\footnote{The motivation for this particular partition was a desire to group based on year of treatment, while also being cognizant of sample sizes. There are five precincts treated between 2012-2014, 12 between 2015-2016, and 11 between 2019-2021.} Despite the various treatment years, each table provides evidence that Cure is associated with a significant reduction in shootings at the time of treatment and afterwards.

Likewise, the geographic variation in treatment makes it unlikely that local shocks could explain our results. As mentioned, Figure \ref{fig:avgShootingsMap} provides a map showing where Cure precincts are located in NYC. While clustered within boroughs, the Cure precincts are scattered throughout all five boroughs. Furthermore, the 2012-2014 treated precincts are even more spread out---they are spread across four boroughs and none of them are contiguous. Despite this geographic variation, Table \ref{tab:earlyCurePoisDIDsTab} shows that Cure is associated with a significant reduction in shootings in these precincts at the exact time of treatment and in the years after.

To further argue against local shocks, we add additional controls to our models. First, we use felony crimes reported to the NYPD to control for location-specific time trends in serious crime.\footnote{Complaints data is public and available \href{https://data.cityofnewyork.us/Public-Safety/NYPD-Complaint-Data-Historic/qgea-i56i/about_data}{here} on NYC Open Data.} Table \ref{tab:mainPoisDIDsComplaintsTab} and Table \ref{tab:mainPoisESComplaintsTab} replicate our main DID and event study models, respectively, while accounting for felony complaints. Relative to our main results, the estimates in these tables suggest that even larger reductions in shootings are associated with Cure treatment, showing our results are robust to crime trends within precincts.

Next, we use felony arrests made by the NYPD to control for location-specific trends in the level of police enforcement of serious crime.\footnote{Arrests data is public and available \href{https://data.cityofnewyork.us/Public-Safety/NYPD-Arrests-Data-Historic-/8h9b-rp9u/about_data}{here} on NYC Open Data.} Table \ref{tab:mainPoisDIDsComplaintsArrestsTab} and Table \ref{tab:mainPoisESComplaintsArrestsTab} add the number of felony arrests to the regressions. The results are practically identical to those in Table \ref{tab:mainPoisDIDsComplaintsTab} and Table \ref{tab:mainPoisESComplaintsTab}, showing our results are robust to changes in NYPD enforcement within precincts. In analyses not shown, we used precinct-year data on NYPD funding and officer headcount as alternative proxies for NYPD enforcement, and our results are robust to both of these proxies. Additionally, we note that felony complaints and arrests explain variation in shootings in our Cure-only sample, resulting in Cure treatment being associated with significant and larger reductions in shootings relative to Table \ref{tab:onlyCurePoisDIDsTab} (see Table \ref{tab:onlyCurePoisDIDsComplaintsTab} and Table \ref{tab:onlyCurePoisDIDsComplaintsArrestsTab}), strengthening the argument against a common time trend within Cure precincts. Beyond our own analyses, we note that previous studies in NYC and elsewhere have accounted for various measures of police activity and enforcement and found that they were unable to explain the reduction in violence associated with Cure Violence (\cite{ButtsEtAl15b}; \cite{HenryEtAl14}; \cite{PicardFritscheCerniglia13}; \cite{WebsterEtAl12}).\footnote{\textcite{PicardFritscheCerniglia13} note that various policing strategies have been used to reduce gun violence in large cities. While some strategies have shown promise, the estimated effect sizes have been relatively small.}

Finally, based on discussions we have had within the City Council and with other City officials, we know of no other programs nor changes outside of CMS that can explain our results.

\subsection*{Spillover Effects}
\subsubsection*{Evidence for Spillover Effects}

As can be seen in Figure \ref{fig:avgShootingsMap}, Cure precincts are generally clustered within boroughs. Gun violence partly operates within social networks---many shootings are in retaliation to earlier shootings or acts of violence (\cite{ButtsEtAl15a})---and social networks tend to form with those nearby. Cure program participants are free to move about and, thus, "treating" a participant in one precinct may prevent a shooting in another precinct. Cure program workers are also free to intervene in conflicts outside their catchment area---\textcite{SkoganEtAl09} found that only half of surveyed Cure participants resided in or frequented a Cure catchment area and Cure staff conducted work over a broad area.\footnote{Anecdotally, a Cure worker in Brownsville told us that they often hear of conflicts brewing outside of their catchment area and will intervene if they have the manpower to do so.} Indeed, this potential for spillover has been commented on in previous Cure research and estimated in one study (e.g., \cite{SkoganEtAl09}; \cite{WebsterEtAl12}).

Figure \ref{fig:numShootings_byYearCure} shows that significant decreases in Cure precincts' shootings began around 2012, even for precincts that had not been treated yet. We have argued that this is unlikely to be due to something other than Cure, however spillover effects could explain the time series we see in Figure \ref{fig:numShootings_byYearCure}.

To assess potential spillover effects, we use the framework described in Section 3. In particular, we say that any precinct, including treated precincts, sharing a border with a Cure precinct is a ``spillover precinct,'' and we define such a precinct's ``spillover treatment date'' as the earliest year in which a contiguous precinct received Cure. With this framework, we can investigate the possibility of ``spillover effects.''  We utilize analogous regression models to those used to estimate Cure effects, but with the treatment and treatment date now referring to potential Cure spillover.

Table \ref{tab:SOBeforeCurePoisSODIDsTab} presents DID analysis of spillover effects. In this table, the control group is control precincts that could not have been spilled into (i.e., do not border a Cure precinct) and the treated group is control and Cure precincts that could have been spilled into.\footnote{As we are interested in measuring spillover effects on ``untreated'' precincts, the table excludes Cure precincts that received Cure before their spillover treatment date.} Using only a binary spillover treatment variable, the first column reports a marginally significant 7.8\% decrease in shootings associated with spillover ``treatment.'' Allowing for dynamic effects, the second and third columns show significant further decreases with time. Additionally, Table \ref{tab:SOBeforeCurePoisSODIDsComplaintsTab} and Table \ref{tab:SOBeforeCurePoisSODIDsComplaintsArrestsTab} show that controlling for felony complaints and arrests results in significant, larger estimates of spillover effects, relative to Table \ref{tab:SOBeforeCurePoisSODIDsTab}.

\begin{figure}[!htbp]
    \centering
    \caption{Number of Shootings Over Time: Control Precincts}\label{fig:numShootings_byControlSO}
    \includegraphics[width=\textwidth]{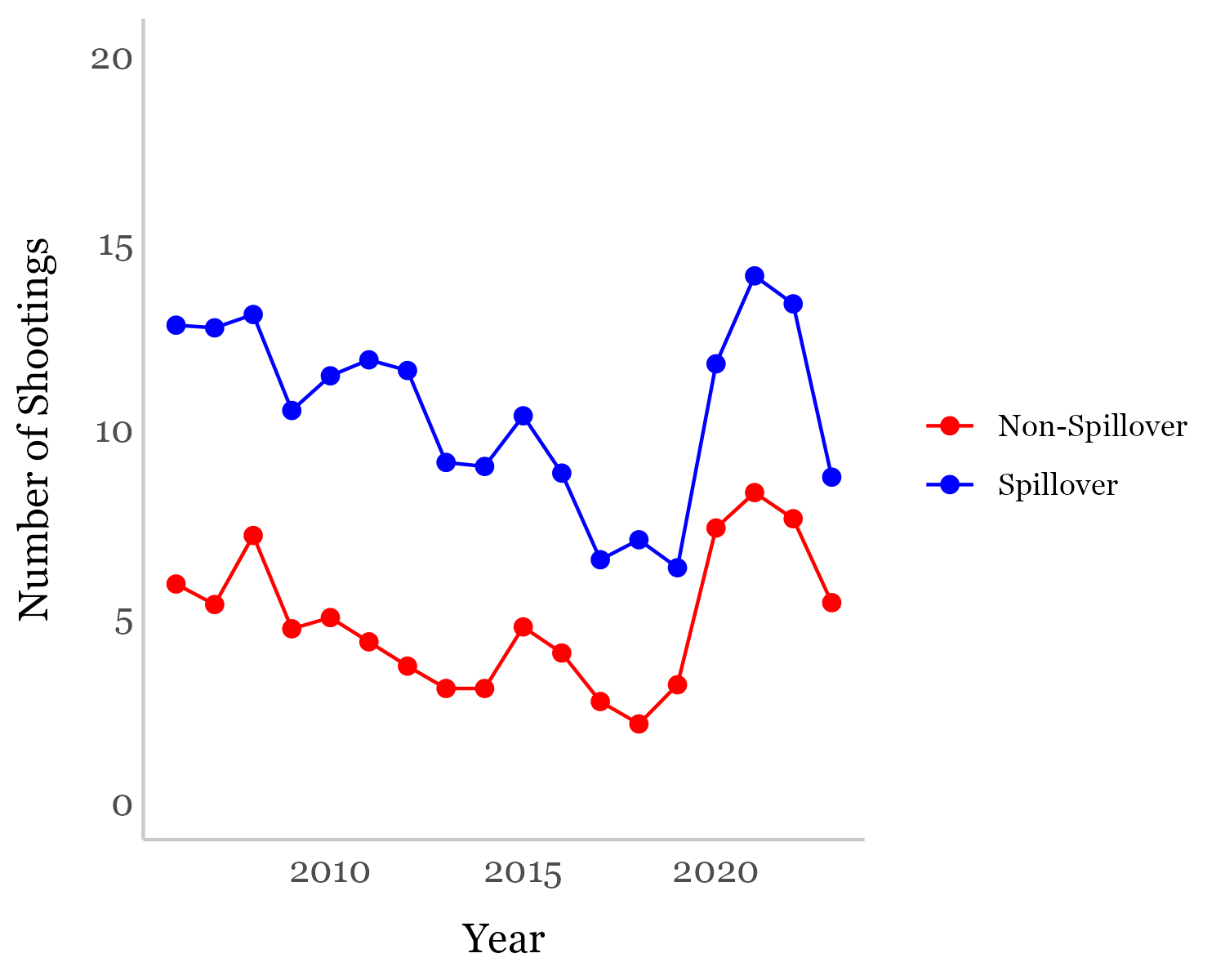}
    \raggedright
    \small{\textit{Note}: This figure plots the average number of shootings (precinct-level) over time for control precincts that could not and could have been spilled over into (according to the spillover assumption in Section 3). This figure excludes all Cure precincts.}
\end{figure}

\begin{figure}[!htbp]
    \centering
    \caption{Spillover Effects: Only Control Precincts}\label{fig:onlyControlsPoisSOES}
    \includegraphics[width=\textwidth]{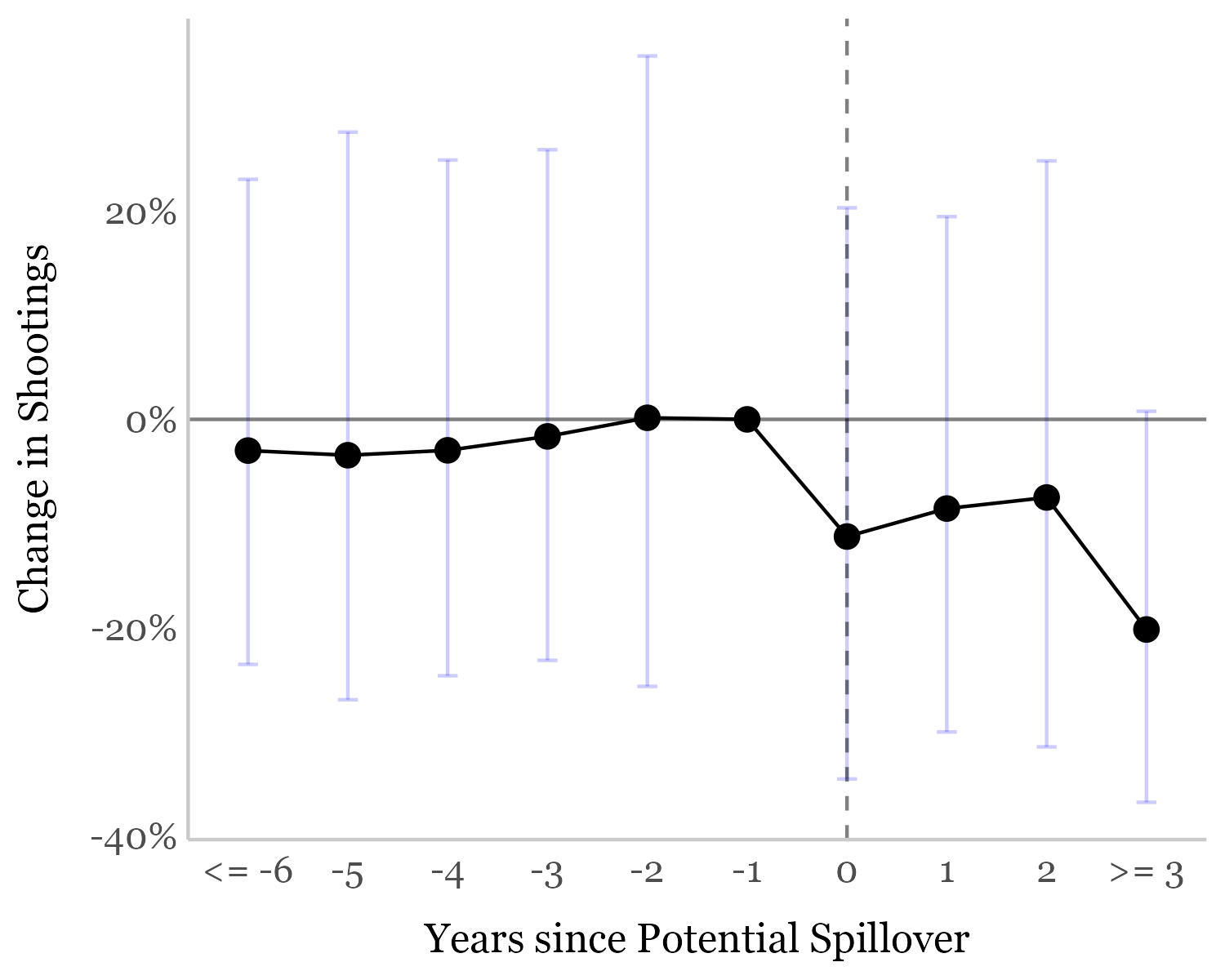}
    \raggedright
    \small{\textit{Note}: This figure plots spillover event study ``treatment'' coefficients, using the setup described in Section 3. The first year pre-spillover is the reference period. This figure excludes all Cure precincts. 95\% confidence intervals are shown.}
\end{figure}

The treatment group in Table \ref{tab:SOBeforeCurePoisSODIDsTab} includes Cure precincts, so the significant long-term effects seen in Table \ref{tab:SOBeforeCurePoisSODIDsTab} may just be the result of receiving Cure in the future. To rule this out, Table \ref{tab:onlyControlsPoisSODIDsTab} presents analogous DID results that exclude all Cure precincts---i.e., the treatment group is control precincts that could have been spilled into. The results in this table reaffirm those in Table \ref{tab:SOBeforeCurePoisSODIDsTab}. The binary spillover treatment is associated with a significant 13.1\% decrease in shootings. The second column makes clear that this results from a small, insignificant initial reduction in shootings that significantly grows with time---perhaps because other nearby Cure precincts receive Cure over time.\footnote{While the results are noisy and the estimates in the third column are not significant, they also suggest a small initial spillover effect that grows with time. Table \ref{tab:onlyControlsPoisSODIDsComplaintsTab} and Table \ref{tab:onlyControlsPoisSODIDsComplaintsArrestsTab} show that controlling for felony complaints and arrests results in very similar estimates to those in Table \ref{tab:onlyControlsPoisSODIDsTab}, but with larger immediate/base effects.} Figure \ref{fig:onlyControlsPoisSOES} presents event study results.\footnote{Figure \ref{fig:numShootings_byControlSO} plots the time series for the average number of shootings for the control precincts that could and could not have been spilled into. The spillover precincts have a higher number of shootings on average and the two types of control precincts follow similar time trends, with a slightly steeper decline in shootings for the spillover precincts.} The specification is demanding and the sample is reduced to only control units, so the estimates are noisy. However, the event study illustrates a flat, insignificant pre-trend and a clear drop in shootings at \emph{exactly} the time of spillover ``treatment,'' which grows and becomes marginally significant in the long-run. Table \ref{tab:onlyControlsPoisSODIDsTab} and Figure \ref{fig:onlyControlsPoisSOES} are striking because they solely use precincts which never received Cure, thus presenting clear suggestive evidence for Cure spillover effects.

\subsubsection*{Implications of Spillover Effects}

As mentioned, spillover effects imply a violation of SUTVA. If nearby precincts are being partially treated by spillovers then they can no longer serve as control units for Cure precincts, and the treatment effect estimates in Table \ref{tab:mainPoisDIDsTab} and Figure \ref{fig:mainPoisES} may be biased. With that said, spillover effects would imply attenuation bias---i.e., our estimates may be a lower bound on Cure's effectiveness at reducing gun violence.

\begin{figure}[!htbp]
    \centering
    \caption{Impact of Cure: No Control Spillover}\label{fig:NonSOControlsPoisES}
    \includegraphics[width=\textwidth]{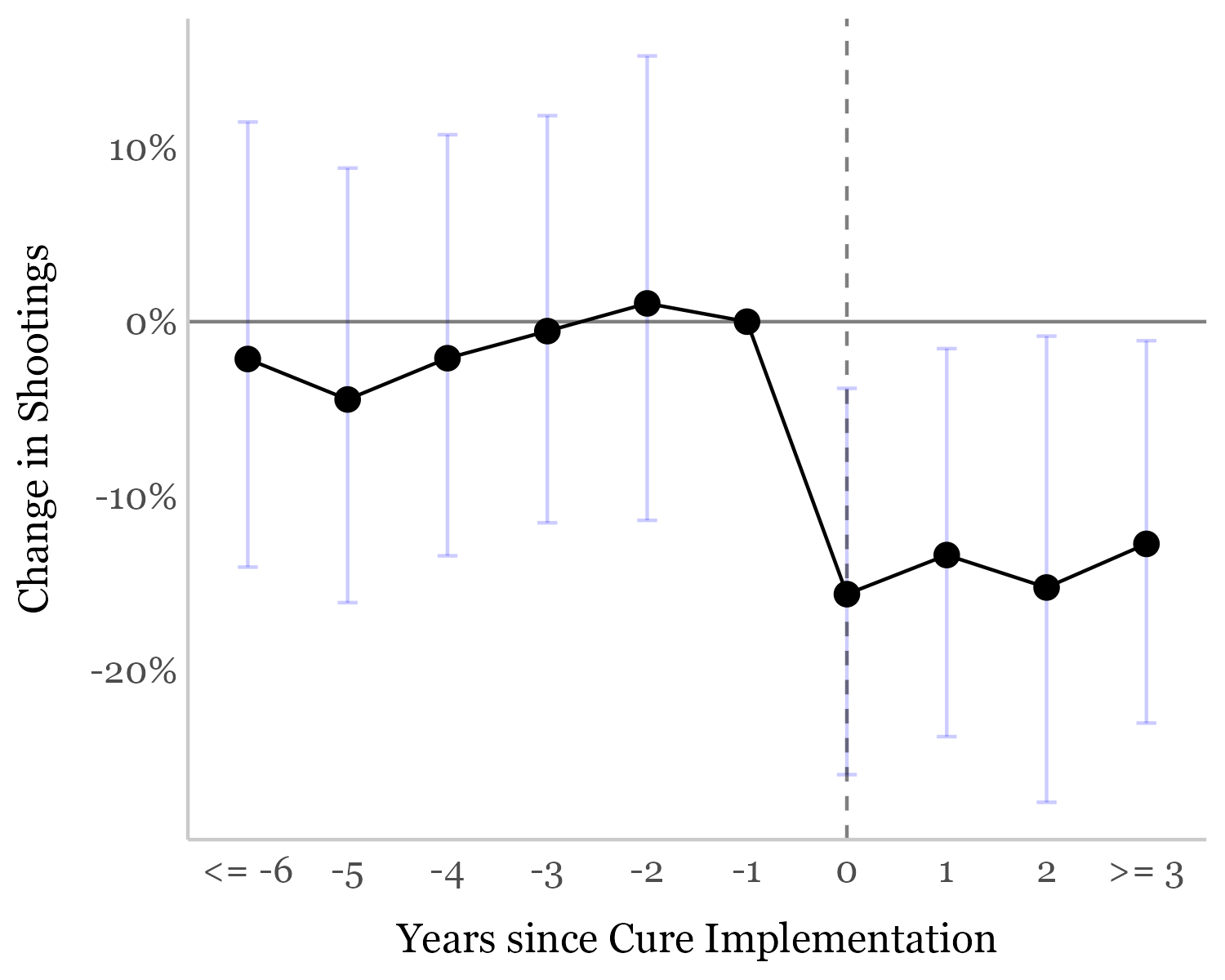}
    \raggedright
    \small{\textit{Note}: This figure plots event study treatment coefficients, using the setup described in Section 3. The first year pre-Cure is the reference period. This figure excludes control precincts that could have been spilled over into. 95\% confidence intervals are shown.}
\end{figure}

As discussed, Table \ref{tab:onlyControlsPoisSODIDsTab} and Figure \ref{fig:onlyControlsPoisSOES} present evidence that there are spillover effects from Cure precincts to nearby control precincts. Thus, we can try to limit bias by excluding these problematic control precincts from our analysis. Table \ref{tab:NonSOControlsPoisDIDsTab} and Figure \ref{fig:NonSOControlsPoisES} present Cure treatment DID and event study results, respectively, using a sample that excludes all control precincts that could have been spilled into. One can see that Table \ref{tab:NonSOControlsPoisDIDsTab} and Figure \ref{fig:NonSOControlsPoisES} are practically identical to our main results (Table \ref{tab:mainPoisDIDsTab} and Figure \ref{fig:mainPoisES}), providing some reassurance that our results are not meaningfully biased by any spillover to control precincts.

There may still be bias caused by spillover between Cure precincts. Table \ref{tab:NonSOControls_NonSOCurePoisDIDsTab} attempts to remove this bias by using only control and Cure precincts that could \emph{not} have been spilled into. The estimates in Table \ref{tab:NonSOControls_NonSOCurePoisDIDsTab} once again suggest that Cure significantly reduces shootings. In fact, the estimates are much larger in magnitude than those in Table \ref{tab:mainPoisDIDsTab}, possibly indicating that our main results are attenuated by spillover effects between Cure precincts. Unfortunately, there are only four Cure precincts that do not share a border with another Cure precinct. Thus, the treated sample in Table \ref{tab:NonSOControls_NonSOCurePoisDIDsTab} is small and possibly non-representative and the estimates in the table are noisy, so the reader should interpret this table cautiously. 

Aside from treatment effect estimation, spillover effects also raise an important policy question. In particular, if Cure treatment is spilling over into nearby precincts, then there may be diminishing returns to implementing new Cure programs in those precincts. On the other hand, Cure programs target small catchment areas within precincts, and it is not obvious whether nearby Cure programs would ``crowd each other out.'' Whether there are decreasing returns to scale or not is critical for policymakers to understand as they determine the best path forward for the city's Cure Violence programs.

To provide some evidence on this question, Table \ref{tab:NonSOControls_SOBeforeCurePoisDIDsTab} presents Cure treatment DID results using a sample that excludes all Cure precincts except those that could have been spilled into before their Cure treatment.\footnote{Table \ref{tab:NonSOControls_SOBeforeCurePoisDIDsTab} also excludes control precincts that could have been spilled into---inclusion/exclusion of these precincts does not affect the results.} Even though we are only utilizing potentially ``post-spillover'' Cure precincts, the table still indicates that Cure treatment is associated with a significant reduction in shootings. The binary treatment variable suggests a significant 18\% reduction in shootings, but there also appear to be meaningful dynamic effects---the quadratic specification suggests a significant base reduction of 27.7\% that wanes in the mid-term and grows in the long-term.\footnote{Table \ref{tab:NonSOControls_SOBeforeCurePoisDIDsComplaintsTab} and Table \ref{tab:NonSOControls_SOBeforeCurePoisDIDsComplaintsArrestsTab} show that controlling for felony complaints and arrests results in very similar estimates to those in Table \ref{tab:NonSOControls_SOBeforeCurePoisDIDsTab}.}

Table \ref{tab:NonSOControls_SOBeforeCurePoisDIDsTab} shows no signs of diminishing returns to opening new Cure programs in precincts nearby existing programs. If anything, the results suggest \emph{increasing} returns to scale, as the estimated Cure effects are larger than those in Table \ref{tab:mainPoisDIDsTab}. However, we caution that the set of ``post-spillover'' Cure precincts includes those treated between 2019-2021, so the Covid pandemic may be influencing the results in Table \ref{tab:NonSOControls_SOBeforeCurePoisDIDsTab}.\footnote{Indeed, comparing Table \ref{tab:earlyCurePoisDIDsTab}, Table \ref{tab:midCurePoisDIDsTab}, and Table \ref{tab:lateCurePoisDIDsTab} shows that estimated Cure effects are largest for Cure precincts treated between 2019-2021 and smallest for those treated between 2015-2016---the two groups from which the ``post-spillover'' Cure precincts come from. Regardless, all three tables find that Cure is significantly associated with reductions in shootings.} 

To further investigate the impact of expanding Cure, we can also check for spillover effects into precincts that already received Cure. Table \ref{tab:NonSOControls_SOAfterCurePoisSODIDsTab} presents spillover DID results that uses a sample that excludes Cure precincts that could have been spilled into before their Cure treatment and excludes control precincts that could have been spilled into. In column one, the binary spillover treatment variable is associated with a significant 22.4\% reduction in shootings. The other columns indicate that this reduction in shootings occurs at the \emph{exact} time of spillover ``treatment'' and does not wane nor grow with time. Given this alignment with the time of potential spillover, the table strongly suggests that opening new Cure programs can reduce shootings in adjacent precincts that already have Cure.\footnote{Table \ref{tab:NonSOControls_SOAfterCurePoisSODIDsComplaintsTab} and Table \ref{tab:NonSOControls_SOAfterCurePoisSODIDsComplaintsArrestsTab} show that controlling for felony complaints and arrests results in very similar, but slightly larger estimates than those in Table \ref{tab:NonSOControls_SOAfterCurePoisSODIDsTab}.} Additionally, the estimated reductions in Table \ref{tab:NonSOControls_SOAfterCurePoisSODIDsTab} are even larger than the Cure effects estimated in Table \ref{tab:mainPoisDIDsTab}. All of this suggests that there are increasing returns to scale to opening new Cure programs.\footnote{Past research hints at why this might occur. \textcite{ButtsEtAl15a} suggest that collaborations between Cure sites might locate and prevent shootings that would have occurred otherwise. \textcite{WebsterEtAl12} note that there was some staff sharing between Baltimore Cure programs when the need arose.} At the very least, concerns about decreasing returns to scale are not supported by our data.

\section{Cost-Benefit Analysis}
\label{sec:cba}

Past Cure Violence research has not conducted a full cost-benefit analysis to determine if Cure is a cost-effective approach to reducing gun violence.\footnote{In one exception to this, Cure Violence Global contracted with an independent researcher to conduct a cost-benefit analysis for Cure in Chicago. While there are conflict of interest concerns, the researcher concluded that Cure saved \$33 for every \$1 spent (\cite{CVG21}).} This omission has been lamented by policymakers and has hindered the expansion of CVI programs, like Cure Violence (\cite{ButtsEtAl15a}). To help fill this gap in the research, this section calculates a benefit-cost ratio for Cure Violence (or CMS) in NYC.

\subsection*{Cost of Gun Violence}

To approximate the cost of gun violence, we use the estimate from \textcite{LudwigCook01}. Using a nationally representative telephone survey conducted in 1998, \textcite{LudwigCook01} use the contingent valuation method to elicit over 1,200 households' willingness-to-pay (WTP) to reduce gun assaults. Specifically, they asked respondents whether they would vote for or against a new government program that would result in 30\% fewer gun assaults, but would cost an extra \$X in annual taxes.\footnote{To not raise concerns about gun rights, the question made clear that the hypothetical reduction in gun violence would result from reduced criminal gun assaults.} The number X was randomly assigned and a follow-up question helped to bound the respondent's WTP for the program. The authors estimate that the public would be willing to pay \$24.5 billion for such a program, or \$1.2 million per gun assault in 1998. Adjusting for inflation, their estimate suggests each gun assault costs society about \$2.24 million (in 2023 dollars).\footnote{To adjust for inflation, we used the CPI-U (annual average numbers, series ID: CUUR0000SA0).}

One may wonder why the contingent valuation method is used to put a price on gun violence. The leading alternative methodology attempts to sum all of the costs of gun violence. Such an approach requires knowledge of, e.g., medical costs, legal system costs, and the cost of lost productivity due to gun violence. While it may be possible to estimate some of these costs, other less direct costs are difficult or impossible to reliably estimate. Such an approach is likely to understate the true societal cost of gun violence. In contrast, at least in theory, respondents should be considering all direct and indirect costs of gun violence when they provide their WTP.\footnote{See \textcite{LudwigCook01} for further discussion and citations on the theoretical basis for the WTP approach.}

The survey literature has repeatedly shown that respondents are able to and willing to answer hypothetical questions of a probabilistic nature, such as the one asked in \textcite{LudwigCook01}. In particular, responses to such questions exhibit high face validity when the questions pertain to well-defined events that are relevant to respondents' lives (\cite{Manski18}). Most respondents to \textcite{LudwigCook01} likely had some subjective beliefs about the risk of gun violence and, if they had ever voted, experience with referendum-style questions that ask them to weigh a presumably desirable policy against a raise in taxes. Therefore, it is not surprising that \textcite{LudwigCook01} found that respondents provided seemingly valid responses that were broadly consistent with economic theory. For example, respondents' WTP was positively correlated with household income and negatively correlated with gun ownership, and approval for the hypothetical program decreased as the tax increase grew. Additionally, the \textcite{LudwigCook01} estimate appears robust, ranging from \$1.87-\$2.24 million under a range of assumptions.

The \textcite{LudwigCook01} estimate also appears to be externally reasonable and consistent with other estimates. They note that methodologies that add the direct costs of gun violence plus some estimated value of health/life generated a total cost close to theirs---demonstrating that respondents' subjective beliefs about the risk and costs of gun violence appear to be meaningfully related to objective reality. Additionally, the \$2.24 million estimate implies a ``value of statistical life'' of (at most) \$11.8 million, which is close to that used by U.S. government agencies (\cite{EPA24}).\footnote{This calculation assumes that 19\% of gun assaults ends in a homicide, which is what we found in our data. We used the CPI-U (annual average numbers, series ID: CUUR0000SA0) to adjust the EPA's estimate to 2023 dollars.} \textcite{LudwigCook01} found that a 30\% reduction in gun violence was worth \$45.8 billion to society, which seems reasonable given that estimates of the total cost of gun violence in the U.S. range from \$229-\$557 billion (\cite{Gobbo23}). Additionally, given the number of U.S. households in 1998, the \$45.8 billion amounts to the average household being willing to pay \$447 dollars more in annual taxes, a modest amount for a 30\% reduction in gun assaults. 

\subsection*{Benefit-Cost Ratio}

We now calculate a benefit-cost ratio for Cure Violence (or CMS). To simplify our calculation, we use the result from the first column of Table \ref{tab:mainPoisDIDsTab}---i.e., on average, Cure led to a 14\% reduction in shootings relative to the counterfactual. The third column of Table \ref{tab:mainPoisDIDsTab} suggests that the treatment effect does not meaningfully change with time, so we use this average 14\% gap in each year post-Cure. We note that using this estimate ignores any spillover to control precincts (Table \ref{tab:onlyControlsPoisSODIDsTab}), and this may result in an underestimate of the benefits of Cure.

Between 2012-2023, police precincts with an active Cure program averaged 40.2 shootings per year. Given this, our DID estimate implies that these precincts would have averaged 46.7 shootings per year in the absence of Cure. Between 2012-2023, in an average year, there were 16.6 precincts with active Cure programs and 1,415 shootings per year. Our estimate implies that, on average, NYC would have seen 108 or 7.6\% more shootings per year between 2012-2023 in the absence of Cure. Over those years, this adds up to 1,296 shootings avoided and---using the \textcite{LudwigCook01} estimate---this translates to \$2.9 billion in social welfare gained. Using CMS historical budget information provided to us by NYC's Office of Management and Budget, the total cost of CMS from FY2013-FY2023 was about \$450 million.\footnote{We use the entire CMS budget, rather than the portion dedicated to Cure, because we are not able to separate out Cure's impact from CMS and because this choice stacks the cards against reaching a net surplus. If most of the benefits of CMS are the result of Cure alone, then our benefit-cost ratio will be an underestimate.} Therefore, Cure Violence in NYC has thus far generated a net social surplus of \$2.45 billion, with a historical benefit-cost ratio of about 6.5:1.

Due to the spike in gun violence after the pandemic, NYC has invested significantly more in Cure Violence and CMS recently, with the FY2023 budget about 2.5 times the FY2020 budget. The relevant question facing policymakers today is whether this larger budget is justified. There were 28 precincts with Cure programs in 2023 and, on average, each Cure precinct witnessed 32.1 shootings in that year. In the absence of Cure, we would have expected Cure precincts to have averaged 37.3 shootings in 2023. This adds up to 145 shootings avoided or 11.6\% of the 1,250 shootings in NYC in 2023. Avoiding these shootings generated social welfare of \$325 million and CMS's budget was \$103 million. Therefore, even with the larger budget, Cure Violence still recently generated a net social surplus of \$222 million, with a benefit-cost ratio of more than 3:1.   

As with any cost-benefit analysis, our analysis is not meant to be comprehensive---e.g., we ignored the opportunity cost of CMS's budget (which would raise costs) and we ignored any spillover effects to control precincts (which would raise benefits). Additionally, with a topic such as gun violence, we stress that there is inherent uncertainty in each component of our analysis. Confidence intervals would likely be misleading, given this uncertainty. Instead, we take comfort in the magnitude of our benefit-cost ratio estimate. Taking benefits as given, our 3:1 ratio suggests that CMS's budget would need to triple from its current size---without reaping any additional benefits---before the costs outweigh the benefits. Taking costs as given, the \textcite{LudwigCook01} estimate of the cost of a gun assault would need to drop to \$710,000 before CMS costs outweigh benefits. Alternatively, our estimate of Cure's impact on gun violence would need to be off by a factor of three before CMS costs outweigh benefits---this is highly unlikely, given our standard error in Table \ref{tab:mainPoisDIDsTab}. To put this last point differently, as long as Cure prevents two shootings per year in each precinct in which it operates, it will be more than paying for itself.

\section{Discussion}
\label{sec:conclusion}
\subsection*{Limitations}

There are important limitations to our work that future research should seek to address. First, Cure Violence treatment in NYC was not randomized. With the politics involved and the independent CBOs that manage Cure programs, running an experiment would be challenging. However, given the costs of gun violence and the considerable investments being made in these programs, we urge policymakers and CBOs to collaborate with researchers \emph{early} in order to design implementation plans that allow for causal analysis.

Second, we do not provide evidence on \emph{why} Cure Violence appears to reduce gun violence.\footnote{Additionally, aside from shootings, our work neglects to look at other outcomes that Cure seeks to impact. Some previous research has utilized surveys to assess if Cure affects communities' beliefs/norms surrounding gun violence (see, e.g., \cite{DelgadoEtAl17}; \cite{PicardFritscheCerniglia13}), but this key outcome variable remains understudied.} As mentioned, we do not disentangle the impact of Cure from CMS more broadly. Even within Cure, it is not clear which component has the greatest impact on gun violence. Previous research has hinted at possible mechanisms, but this remains a significant open question in the literature.\footnote{E.g., \textcite{WebsterEtAl12} suggest that the frequency of conflict mediations partially explains variation in Cure's impact on homicides. In simulated models, \textcite{CerdaEtAl18} also found that the addition of more violence interrupters was the critical driver of Cure's impact.} With a better understanding of the mechanisms, policymakers and CBOs could allocate funding and institute ``best practices'' in order to optimize effectiveness. Uncovering these mechanisms lends itself well to experimentation, providing further motivation for policymakers and CBOs to collaborate with researchers.

Third, we do not investigate the efficacy of other gun violence prevention strategies. Likewise, we do not evaluate whether Cure Violence is the most cost-effective gun violence prevention strategy.\footnote{In a computational study, \textcite{CerdaEtAl18} empirically calibrated an agent-based model of the NYC adult population to simulate and estimate the causal effects of Cure Violence and hot spots policing. They found that implementing Cure Violence for a decade reduced violent victimization by 13\%, while implementing hot spots policing and doubling the police force for a decade reduced victimization by 11\%---suggesting that Cure is a more cost-effective approach to violence reduction. However, they also found that a multi-pronged approach, combining Cure with more policing, achieved larger reductions in victimization, with less resources and in a shorter time period, than either approach alone.}

\subsection*{Policy}

Past research and conversations we had with policymakers and domain experts suggest beneficial policies to consider.\footnote{Some policy ideas here stem from a panel we hosted during NYC ``Open Data Week.'' We thank Council Members Althea Stevens and Yusef Salaam, Dr. Jeffrey A. Butts, Director of the Research and Evaluation Center at John Jay College of Criminal Justice, R. Brent Decker, Chief Program Officer of Cure Violence Global, and Hector Cuevas, VP of Education and Youth Development Programs at CAMBA, for taking part in our panel and contributing their thoughts and ideas. While some policy ideas here are targeted towards NYC, we hope other localities can learn from them, as well.} Given that Cure programs operate autonomously, are not an arm of law enforcement, and frequently employ workers with criminal histories, Cure Violence struggles for recognition and acceptance, leading to insufficient and uncertain funding/resources (\cite{PuglieseEtAl22}).

A lack of sufficient, stable funding and resources could limit Cure's ability to reduce gun violence. Low pay, long and unstable hours, and continued exposure to traumatic events can lead to high stress and turnover of Cure staff (\cite{PuglieseEtAl22}). Outreach workers can only handle a caseload of so many high-risk individuals and insufficient staffing will lead to missed opportunities.\footnote{A CVI program in Chicago that similarly targeted high-risk individuals for participation found that for every one program participant there were more than 20 others with similar risk profiles who were not able to receive services (\cite{Northwestern21}). See also \textcite{MaguireEtAl18}.} In Baltimore, \textcite{WebsterEtAl12} noted that Cure workers in one catchment area were redirected to help overwhelmed staff in another area, hindering violence prevention in the staff's original area. In Chicago, \textcite{SkoganEtAl09} found that unstable funding created job uncertainty for Cure workers, drew staff time away from operations, and led to service interruptions.

Luckily, political support for CVI programs has grown since the onset of the pandemic. State and local funding stemming from the 2021 American Rescue Plan can and has been used to start and expand CVI programs, including Cure Violence (\cite{MacGillis23}). Further federal support could be on the way via the Break the Cycle of Violence Act, which would allocate \$5 billion to effective CVI approaches (\cite{OMB22}). Stable federal funding will likely require further evidence of CVI programs' efficacy, requiring more data. While Cure workers are taught to track their work, already stretched bandwidths can prevent high-quality data collection. Policymakers and CBOs should consider additional funding to allow for dedicated data and grant writing staff in order to better secure stable funding.\footnote{Cure workers take program participants' privacy seriously. Thus, data on program participants must be de-identified and policymakers and law enforcement should be respectful of this.}

NYC's CMS contracts dictate exactly which CMS services (e.g., Cure services or wrap-around services) a CBO can provide. This contract system has created confusion and inefficiencies. For example, a CBO with a Cure Violence contract may not have a contract to provide wrap-around services, it may be unclear which CBO has the necessary contract, and CBOs can lose visibility of participants' progress once they are referred to another CBO. City officials should consider using more holistic CMS contracts or creating a centralized system where staff can quickly connect participants with the services that they need. In a related issue, communication between CBOs in the CMS network is limited. This lack of communication partly stems from bandwidth issues, as workers are stretched just to handle day-to-day operations. Communication could be improved by implementing a centralized communication system designed to collect and distribute pertinent information, allowing Cure workers to quickly react to changes in the broader environment.

Cure Violence program placement and funding should be determined based on need. With that said, successful Cure programs require competent CBOs to manage them.\footnote{\textcite{MaguireEtAl18} note that poor management negatively affected day-to-day Cure Violence operations.} Established CBOs likely have more stable funding, as well as a greater number of staff in place to handle administrative tasks (\cite{SkoganEtAl09}). However, the areas that could benefit from Cure Violence do not necessarily overlap with the location of high-functioning CBOs. In such cases, the City should provide local CBOs with additional support, so that Cure workers can focus on their core responsibilities rather than administrative functions.

Frictions can arise between law enforcement and Cure Violence programs. For instance, in the past year, NYPD officers have cursed at, shoved, and arrested some Cure workers for minor offenses (\cite{CramerMeko24}). Anecdotally, we have been told that it takes time for Cure programs to establish good relationships with police, but this work must start from scratch whenever there is turnover in police personnel. Police academy training should teach cadets about Cure Violence and other CVIs and the value of the work that they do. Officers should understand that Cure workers do not work for or with the police and should learn how to appropriately work alongside Cure workers.

\subsection*{Conclusion}

Cure Violence is a CVI program that aims to reduce gun violence by mediating conflicts, ``treating'' high-risk individuals, and changing community norms. Using NYC shootings data from 2006-2023, with 28 Cure and 48 control police precincts, we assess the efficacy of Cure Violence using both DID and event study models. We find that, on average, Cure Violence is associated with a 14\% reduction in shootings relative to the counterfactual. This association persists in the years after treatment, neither increasing nor decreasing much over time. We exploit variation in the geography and timing of Cure implementation to argue against alternative explanations. Additionally, we find suggestive evidence of spillover effects into nearby precincts, as well as increasing returns to opening new Cure programs. Interpreted causally, our results imply that around 1,300 shootings were avoided between 2012-2023 due to Cure---generating a net social surplus of \$2.45 billion and achieving a benefit-cost ratio of 6.5:1.

With a limited budget, policymakers must assess the evidence for Cure Violence and other violence reduction strategies, and decide how best to allocate resources. While we note caveats to our work, our results strongly suggest that Cure Violence can be an effective \emph{component} of a collection of strategies to combat gun violence. This suggests that further investment in Cure Violence is likely to generate a net social surplus and expanded use of the program may bring further benefits.\footnote{E.g., \textcite{RansfordEtAl16} provide promising evidence that the Cure approach can effectively work in prisons, as well.}

\newpage

\printbibliography
\newpage
\section*{Figures}

\newpage
\section*{Tables}

\begin{table}[!htbp] \centering 
  \caption{Impact of Cure DID Results} 
  \label{tab:mainPoisDIDsTab} 
\begin{tabular}{@{\extracolsep{5pt}}lD{.}{.}{-3} D{.}{.}{-3} D{.}{.}{-3} } 
\\[-1.8ex]\hline 
\hline \\[-1.8ex] 
\\[-1.8ex] & \multicolumn{3}{c}{Log Change in Shootings} \\ 
\\[-1.8ex] & \multicolumn{1}{c}{(1)} & \multicolumn{1}{c}{(2)} & \multicolumn{1}{c}{(3)}\\ 
\hline \\[-1.8ex] 
 Cure Implemented & -0.151^{***} & -0.126^{***} & -0.176^{***} \\ 
  & (0.033) & (0.037) & (0.043) \\ 
  Years of Cure &  & -0.010 & 0.031 \\ 
  &  & (0.007) & (0.020) \\ 
  Years of Cure Sq. &  &  & -0.005^{**} \\ 
  &  &  & (0.002) \\ 
 Observations & \multicolumn{1}{c}{1,368} & \multicolumn{1}{c}{1,368} & \multicolumn{1}{c}{1,368} \\ 
\hline \\[-1.8ex] 
\end{tabular}

\raggedright
\small{\textit{Note}: This table reports treatment coefficients for several DID models, using the setup described in Section 3. ``Cure Implemented'' is an indicator that is on once a precinct is treated. ``Num. Years of Cure'' is a variable for the number of years post-Cure in a precinct---in the third row we add a squared term for this variable. All models include precinct and year fixed effects. Robust standard errors are in parentheses. * p<0.1, ** p<0.05, *** p<0.01.}
\end{table}

\begin{table}[!htbp] \centering 
  \caption{Impact of Cure: Only Cure Precincts}
  \label{tab:onlyCurePoisDIDsTab} 
\begin{tabular}{@{\extracolsep{5pt}}lD{.}{.}{-3} D{.}{.}{-3} D{.}{.}{-3} } 
\\[-1.8ex]\hline 
\hline \\[-1.8ex] 
\\[-1.8ex] & \multicolumn{3}{c}{Log Change in Shootings} \\ 
\\[-1.8ex] & \multicolumn{1}{c}{(1)} & \multicolumn{1}{c}{(2)} & \multicolumn{1}{c}{(3)}\\ 
\hline \\[-1.8ex] 
 Cure Implemented & -0.068^{*} & -0.069^{*} & -0.110^{**} \\ 
  & (0.041) & (0.041) & (0.048) \\ 
  Years of Cure &  & 0.001 & 0.035^{*} \\ 
  &  & (0.009) & (0.021) \\ 
  Years of Cure Sq. &  &  & -0.004^{*} \\ 
  &  &  & (0.002) \\
 Observations & \multicolumn{1}{c}{504} & \multicolumn{1}{c}{504} & \multicolumn{1}{c}{504} \\ 
\hline \\[-1.8ex] 
\end{tabular}

\raggedright
\small{\textit{Note}: This table reports treatment coefficients for several DID models, using the setup described in Section 3. This table excludes all control precincts. ``Cure Implemented'' is an indicator that is on once a precinct is treated. ``Num. Years of Cure'' is a variable for the number of years post-Cure in a precinct---in the third row we add a squared term for this variable. All models include precinct and year fixed effects. Robust standard errors are in parentheses. * p<0.1, ** p<0.05, *** p<0.01.}
\end{table}

\begin{table}[!htbp] \centering 
  \caption{Spillover Effects}
  \label{tab:SOBeforeCurePoisSODIDsTab}
\begin{tabular}{@{\extracolsep{5pt}}lD{.}{.}{-3} D{.}{.}{-3} D{.}{.}{-3} } 
\\[-1.8ex]\hline 
\hline \\[-1.8ex] 
\\[-1.8ex] & \multicolumn{3}{c}{Log Change in Shootings} \\ 
\\[-1.8ex] & \multicolumn{1}{c}{(1)} & \multicolumn{1}{c}{(2)} & \multicolumn{1}{c}{(3)}\\ 
\hline \\[-1.8ex] 
 Spillover Possible & -0.081^{*} & 0.016 & -0.088^{*} \\ 
  & (0.047) & (0.047) & (0.053) \\ 
  Years of Spillover &  & -0.057^{***} & 0.031 \\ 
  &  & (0.009) & (0.024) \\ 
  Years of Spillover Sq. &  &  & -0.009^{***} \\ 
  &  &  & (0.002) \\
 Observations & \multicolumn{1}{c}{1,134} & \multicolumn{1}{c}{1,134} & \multicolumn{1}{c}{1,134} \\ 
\hline \\[-1.8ex] 
\end{tabular}    

\raggedright
\small{\textit{Note}: This table reports spillover ``treatment'' coefficients for several DID models, using the setup described in Section 3. This table excludes Cure precincts with a spillover ``treatment'' date after their Cure treatment. ``Spillover Possible'' is an indicator that is on once a precinct is ``treated'' (i.e., potentially spilled over into). ``Years of Spillover'' is a variable for the number of years post (potential) spillover in a precinct---in the third row we add a squared term for this variable. All models include precinct and year fixed effects. Robust standard errors are in parentheses. * p<0.1, ** p<0.05, *** p<0.01.}
\end{table}

\begin{table}[!htbp] \centering 
  \caption{Spillover Effects: Only Control Precincts}
  \label{tab:onlyControlsPoisSODIDsTab}
\begin{tabular}{@{\extracolsep{5pt}}lD{.}{.}{-3} D{.}{.}{-3} D{.}{.}{-3} } 
\\[-1.8ex]\hline 
\hline \\[-1.8ex] 
\\[-1.8ex] & \multicolumn{3}{c}{Log Change in Shootings} \\ 
\\[-1.8ex] & \multicolumn{1}{c}{(1)} & \multicolumn{1}{c}{(2)} & \multicolumn{1}{c}{(3)}\\ 
\hline \\[-1.8ex] 
 Spillover Possible & -0.140^{**} & -0.033 & -0.067 \\ 
  & (0.067) & (0.072) & (0.087) \\ 
  Years of Spillover &  & -0.040^{***} & -0.014 \\ 
  &  & (0.013) & (0.038) \\ 
  Years of Spillover Sq. &  &  & -0.003 \\ 
  &  &  & (0.004) \\
 Observations & \multicolumn{1}{c}{864} & \multicolumn{1}{c}{864} & \multicolumn{1}{c}{864} \\ 
\hline \\[-1.8ex] 
\end{tabular}     

\raggedright
\small{\textit{Note}: This table reports spillover ``treatment'' coefficients for several DID models, using the setup described in Section 3. This table excludes all Cure precincts. ``Spillover Possible'' is an indicator that is on once a precinct is ``treated'' (i.e., potentially spilled over into). ``Years of Spillover'' is a variable for the number of years post (potential) spillover in a precinct---in the third row we add a squared term for this variable. All models include precinct and year fixed effects. Robust standard errors are in parentheses. * p<0.1, ** p<0.05, *** p<0.01.}
\end{table}

\begin{table}[!htbp] \centering 
  \caption{Impact of Cure on Post-Spillover Precincts}
  \label{tab:NonSOControls_SOBeforeCurePoisDIDsTab}
\begin{tabular}{@{\extracolsep{5pt}}lD{.}{.}{-3} D{.}{.}{-3} D{.}{.}{-3} } 
\\[-1.8ex]\hline 
\hline \\[-1.8ex] 
\\[-1.8ex] & \multicolumn{3}{c}{Log Change in Shootings} \\ 
\\[-1.8ex] & \multicolumn{1}{c}{(1)} & \multicolumn{1}{c}{(2)} & \multicolumn{1}{c}{(3)}\\ 
\hline \\[-1.8ex] 
 Cure Implemented & -0.199^{***} & -0.168^{***} & -0.324^{***} \\ 
  & (0.056) & (0.060) & (0.070) \\ 
  Years of Cure &  & -0.018 & 0.137^{***} \\ 
  &  & (0.014) & (0.042) \\ 
  Years of Cure Sq. &  &  & -0.022^{***} \\ 
  &  &  & (0.006) \\ 
 Observations & \multicolumn{1}{c}{630} & \multicolumn{1}{c}{630} & \multicolumn{1}{c}{630} \\ 
\hline \\[-1.8ex] 
\end{tabular}   

\raggedright
\small{\textit{Note}: This table reports treatment coefficients for several DID models, using the setup described in Section 3. This table excludes Cure precincts except those that could have been spilled over into before their Cure treatment and it excludes control precincts that could have been spilled over into. ``Cure Implemented'' is an indicator that is on once a precinct is treated. ``Num. Years of Cure'' is a variable for the number of years post-Cure in a precinct---in the third row we add a squared term for this variable. All models include precinct and year fixed effects. Robust standard errors are in parentheses. * p<0.1, ** p<0.05, *** p<0.01.}
\end{table}

\begin{table}[!htbp] \centering 
  \caption{Spillover Effects on Already Treated Precincts}
  \label{tab:NonSOControls_SOAfterCurePoisSODIDsTab}
\begin{tabular}{@{\extracolsep{5pt}}lD{.}{.}{-3} D{.}{.}{-3} D{.}{.}{-3} } 
\\[-1.8ex]\hline 
\hline \\[-1.8ex] 
\\[-1.8ex] & \multicolumn{3}{c}{Log Change in Shootings} \\ 
\\[-1.8ex] & \multicolumn{1}{c}{(1)} & \multicolumn{1}{c}{(2)} & \multicolumn{1}{c}{(3)}\\ 
\hline \\[-1.8ex] 
 Spillover Possible & -0.253^{***} & -0.217^{***} & -0.289^{***} \\ 
  & (0.073) & (0.081) & (0.098) \\ 
  Years of Spillover &  & -0.016 & 0.054 \\ 
  &  & (0.016) & (0.050) \\ 
  Years of Spillover Sq. &  &  & -0.009 \\ 
  &  &  & (0.006) \\ 
 Observations & \multicolumn{1}{c}{486} & \multicolumn{1}{c}{486} & \multicolumn{1}{c}{486} \\ 
\hline \\[-1.8ex] 
\end{tabular}       

\raggedright
\small{\textit{Note}: This table reports spillover ``treatment'' coefficients for several DID models, using the setup described in Section 3. This table excludes Cure precincts with a spillover ``treatment'' date before their Cure treatment and it excludes control precincts that could have been spilled over into. ``Spillover Possible'' is an indicator that is on once a precinct is ``treated'' (i.e., potentially spilled over into). ``Years of Spillover'' is a variable for the number of years post (potential) spillover in a precinct---in the third row we add a squared term for this variable. All models include precinct and year fixed effects. Robust standard errors are in parentheses. * p<0.1, ** p<0.05, *** p<0.01.}
\end{table}

\newpage
\appendix
\counterwithin{figure}{section}
\numberwithin{table}{section}
\section{Appendix: Additional Figures}

\newpage
\section{Appendix: Additional Tables}

\begin{table}[!htbp]
    \centering
    \caption{Balance Table}\label{tab:balanceTable}
    \includegraphics[width=\textwidth]{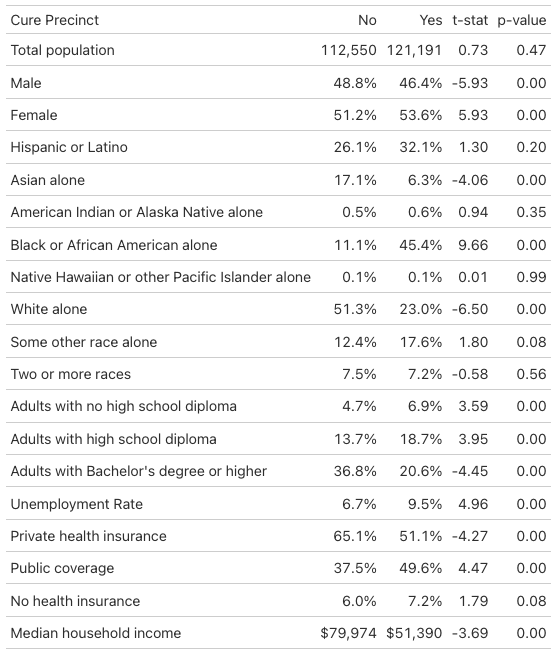}
    \raggedright
    \small{\textit{Note}: This is a balance table between Cure and control precincts. The data is taken from the 2017-2021 American Community Survey and is based off of the authors' calculations to match Census tracts to NYC police precincts.}
\end{table}

\begin{table}[!htbp] \centering 
  \caption{Impact of Cure Event Study} 
  \label{tab:mainPoisESTab} 
\begin{tabular}{@{\extracolsep{5pt}}lD{.}{.}{-3} } 
\\[-1.8ex]\hline 
\hline \\[-1.8ex] 
\\[-1.8ex] & \multicolumn{1}{c}{Log Change in Shootings} \\ 
\hline \\[-1.8ex] 
 <=6 years pre-Cure & -0.021 \\ 
  & (0.057) \\ 
  5 years pre-Cure & -0.045 \\ 
  & (0.062) \\ 
  4 years pre-Cure & -0.024 \\ 
  & (0.061) \\ 
  3 years pre-Cure & -0.015 \\ 
  & (0.058) \\ 
  2 years pre-Cure & 0.009 \\ 
  & (0.066) \\ 
  Cure start year & -0.186^{***} \\ 
  & (0.065) \\ 
  1 year post-Cure & -0.159^{**} \\ 
  & (0.064) \\ 
  2 years post-Cure & -0.176^{**} \\ 
  & (0.079) \\ 
  >=3 years post-Cure & -0.149^{***} \\ 
  & (0.057) \\ 
 Observations & \multicolumn{1}{c}{1,368} \\ 
\hline \\[-1.8ex] 
\end{tabular}

\raggedright
\small{\textit{Note}: This table reports event study treatment coefficients, using the setup described in Section 3. The first year pre-Cure is the reference period. The model includes precinct and year fixed effects. Robust standard errors are in parentheses. * p<0.1, ** p<0.05, *** p<0.01.}
\end{table}

\begin{table}[!htbp] \centering 
  \caption{Impact of Cure: Only Pre-Covid Data}
  \label{tab:preCovidPoisDIDsTab} 
\begin{tabular}{@{\extracolsep{5pt}}lD{.}{.}{-3} D{.}{.}{-3} D{.}{.}{-3} } 
\\[-1.8ex]\hline 
\hline \\[-1.8ex] 
\\[-1.8ex] & \multicolumn{3}{c}{Log Change in Shootings} \\ 
\\[-1.8ex] & \multicolumn{1}{c}{(1)} & \multicolumn{1}{c}{(2)} & \multicolumn{1}{c}{(3)}\\ 
\hline \\[-1.8ex] 
 Cure Implemented & -0.142^{***} & -0.113^{**} & -0.111^{**} \\ 
  & (0.042) & (0.050) & (0.053) \\ 
  Years of Cure &  & -0.019 & -0.023 \\ 
  &  & (0.018) & (0.042) \\ 
  Years of Cure Sq. &  &  & 0.001 \\ 
  &  &  & (0.008) \\
 Observations & \multicolumn{1}{c}{910} & \multicolumn{1}{c}{910} & \multicolumn{1}{c}{910} \\ 
\hline \\[-1.8ex] 
\end{tabular}

\raggedright
\small{\textit{Note}: This table reports treatment coefficients for several DID models, using the setup described in Section 3. This table only uses shootings data from 2006-2019 and excludes Cure precincts treated between 2019-2021. ``Cure Implemented'' is an indicator that is on once a precinct is treated. ``Num. Years of Cure'' is a variable for the number of years post-Cure in a precinct---in the third row we add a squared term for this variable. All models include precinct and year fixed effects. Robust standard errors are in parentheses. * p<0.1, ** p<0.05, *** p<0.01.}
\end{table}

\begin{table}[!htbp] \centering 
  \caption{Impact of Cure: Only 2012-2014 Cure Precincts}
  \label{tab:earlyCurePoisDIDsTab}
\begin{tabular}{@{\extracolsep{5pt}}lD{.}{.}{-3} D{.}{.}{-3} D{.}{.}{-3} } 
\\[-1.8ex]\hline 
\hline \\[-1.8ex] 
\\[-1.8ex] & \multicolumn{3}{c}{Log Change in Shootings} \\ 
\\[-1.8ex] & \multicolumn{1}{c}{(1)} & \multicolumn{1}{c}{(2)} & \multicolumn{1}{c}{(3)}\\ 
\hline \\[-1.8ex] 
 Cure Implemented & -0.248^{***} & -0.154^{**} & -0.170^{**} \\ 
  & (0.053) & (0.063) & (0.070) \\ 
  Years of Cure &  & -0.018^{*} & -0.007 \\ 
  &  & (0.009) & (0.032) \\ 
  Years of Cure Sq. &  &  & -0.001 \\ 
  &  &  & (0.003) \\
 Observations & \multicolumn{1}{c}{954} & \multicolumn{1}{c}{954} & \multicolumn{1}{c}{954} \\ 
\hline \\[-1.8ex] 
\end{tabular}

\raggedright
\small{\textit{Note}: This table reports treatment coefficients for several DID models, using the setup described in Section 3. This table excludes Cure precincts, except those that received Cure between 2012-2014. ``Cure Implemented'' is an indicator that is on once a precinct is treated. ``Num. Years of Cure'' is a variable for the number of years post-Cure in a precinct---in the third row we add a squared term for this variable. All models include precinct and year fixed effects. Robust standard errors are in parentheses. * p<0.1, ** p<0.05, *** p<0.01.}
\end{table}

\begin{table}[!htbp] \centering 
  \caption{Impact of Cure: Only 2015-2016 Cure Precincts}
  \label{tab:midCurePoisDIDsTab}
\begin{tabular}{@{\extracolsep{5pt}}lD{.}{.}{-3} D{.}{.}{-3} D{.}{.}{-3} } 
\\[-1.8ex]\hline 
\hline \\[-1.8ex] 
\\[-1.8ex] & \multicolumn{3}{c}{Log Change in Shootings} \\ 
\\[-1.8ex] & \multicolumn{1}{c}{(1)} & \multicolumn{1}{c}{(2)} & \multicolumn{1}{c}{(3)}\\ 
\hline \\[-1.8ex] 
 Cure Implemented & -0.165^{***} & -0.040 & -0.148^{*} \\ 
  & (0.048) & (0.072) & (0.085) \\ 
  Years of Cure &  & -0.032^{**} & 0.078 \\ 
  &  & (0.015) & (0.052) \\ 
  Years of Cure Sq. &  &  & -0.015^{**} \\ 
  &  &  & (0.007) \\ 
 Observations & \multicolumn{1}{c}{1,080} & \multicolumn{1}{c}{1,080} & \multicolumn{1}{c}{1,080} \\ 
\hline \\[-1.8ex] 
\end{tabular} 

\raggedright
\small{\textit{Note}: This table reports treatment coefficients for several DID models, using the setup described in Section 3. This table excludes Cure precincts, except those that received Cure between 2015-2016. ``Cure Implemented'' is an indicator that is on once a precinct is treated. ``Num. Years of Cure'' is a variable for the number of years post-Cure in a precinct---in the third row we add a squared term for this variable. All models include precinct and year fixed effects. Robust standard errors are in parentheses. * p<0.1, ** p<0.05, *** p<0.01.}
\end{table}

\begin{table}[!htbp] \centering 
  \caption{Impact of Cure: Only 2019-2021 Cure Precincts}
  \label{tab:lateCurePoisDIDsTab}
\begin{tabular}{@{\extracolsep{5pt}}lD{.}{.}{-3} D{.}{.}{-3} D{.}{.}{-3} } 
\\[-1.8ex]\hline 
\hline \\[-1.8ex] 
\\[-1.8ex] & \multicolumn{3}{c}{Log Change in Shootings} \\ 
\\[-1.8ex] & \multicolumn{1}{c}{(1)} & \multicolumn{1}{c}{(2)} & \multicolumn{1}{c}{(3)}\\ 
\hline \\[-1.8ex] 
 Cure Implemented & -0.248^{***} & -0.426^{***} & -0.356^{***} \\ 
  & (0.067) & (0.085) & (0.097) \\ 
  Years of Cure &  & 0.133^{***} & -0.012 \\ 
  &  & (0.045) & (0.118) \\ 
  Years of Cure Sq. &  &  & 0.039 \\ 
  &  &  & (0.029) \\ 
 Observations & \multicolumn{1}{c}{1,062} & \multicolumn{1}{c}{1,062} & \multicolumn{1}{c}{1,062} \\ 
\hline \\[-1.8ex] 
\end{tabular} 

\raggedright
\small{\textit{Note}: This table reports treatment coefficients for several DID models, using the setup described in Section 3. This table excludes Cure precincts, except those that received Cure between 2019-2021. ``Cure Implemented'' is an indicator that is on once a precinct is treated. ``Num. Years of Cure'' is a variable for the number of years post-Cure in a precinct---in the third row we add a squared term for this variable. All models include precinct and year fixed effects. Robust standard errors are in parentheses. * p<0.1, ** p<0.05, *** p<0.01.}
\end{table}

\begin{table}[!htbp] \centering 
  \caption{Impact of Cure: No Control Spillover}
  \label{tab:NonSOControlsPoisDIDsTab}
\begin{tabular}{@{\extracolsep{5pt}}lD{.}{.}{-3} D{.}{.}{-3} D{.}{.}{-3} } 
\\[-1.8ex]\hline 
\hline \\[-1.8ex] 
\\[-1.8ex] & \multicolumn{3}{c}{Log Change in Shootings} \\ 
\\[-1.8ex] & \multicolumn{1}{c}{(1)} & \multicolumn{1}{c}{(2)} & \multicolumn{1}{c}{(3)}\\ 
\hline \\[-1.8ex] 
 Cure Implemented & -0.147^{***} & -0.128^{***} & -0.174^{***} \\ 
  & (0.040) & (0.041) & (0.047) \\ 
  Years of Cure &  & -0.013 & 0.024 \\ 
  &  & (0.008) & (0.021) \\ 
  Years of Cure Sq. &  &  & -0.004^{**} \\ 
  &  &  & (0.002) \\
 Observations & \multicolumn{1}{c}{864} & \multicolumn{1}{c}{864} & \multicolumn{1}{c}{864} \\ 
\hline \\[-1.8ex] 
\end{tabular}  

\raggedright
\small{\textit{Note}: This table reports treatment coefficients for several DID models, using the setup described in Section 3. This table excludes control precincts that could have been spilled over into. ``Cure Implemented'' is an indicator that is on once a precinct is treated. ``Num. Years of Cure'' is a variable for the number of years post-Cure in a precinct---in the third row we add a squared term for this variable. All models include precinct and year fixed effects. Robust standard errors are in parentheses. * p<0.1, ** p<0.05, *** p<0.01.}
\end{table}

\begin{table}[!htbp] \centering 
  \caption{Impact of Cure on Non-Spillover Precincts}
  \label{tab:NonSOControls_NonSOCurePoisDIDsTab}
\begin{tabular}{@{\extracolsep{5pt}}lD{.}{.}{-3} D{.}{.}{-3} D{.}{.}{-3} } 
\\[-1.8ex]\hline 
\hline \\[-1.8ex] 
\\[-1.8ex] & \multicolumn{3}{c}{Log Change in Shootings} \\ 
\\[-1.8ex] & \multicolumn{1}{c}{(1)} & \multicolumn{1}{c}{(2)} & \multicolumn{1}{c}{(3)}\\ 
\hline \\[-1.8ex] 
 Cure Implemented & -0.416^{***} & -0.122 & -0.284^{**} \\ 
  & (0.096) & (0.124) & (0.138) \\ 
  Years of Cure &  & -0.073^{***} & 0.076 \\ 
  &  & (0.023) & (0.079) \\ 
  Years of Cure Sq. &  &  & -0.019^{*} \\ 
  &  &  & (0.010) \\
 Observations & \multicolumn{1}{c}{432} & \multicolumn{1}{c}{432} & \multicolumn{1}{c}{432} \\ 
\hline \\[-1.8ex] 
\end{tabular}    

\raggedright
\small{\textit{Note}: This table reports treatment coefficients for several DID models, using the setup described in Section 3. This table excludes Cure and control precincts that could have been spilled over into. ``Cure Implemented'' is an indicator that is on once a precinct is treated. ``Num. Years of Cure'' is a variable for the number of years post-Cure in a precinct---in the third row we add a squared term for this variable. All models include precinct and year fixed effects. Robust standard errors are in parentheses. * p<0.1, ** p<0.05, *** p<0.01.}
\end{table}

\begin{table}[!htbp] \centering 
  \caption{Impact of Cure DID Results} 
  \label{tab:mainPoisDIDsComplaintsTab} 
\begin{tabular}{@{\extracolsep{5pt}}lD{.}{.}{-3} D{.}{.}{-3} D{.}{.}{-3} } 
\\[-1.8ex]\hline 
\hline \\[-1.8ex] 
\\[-1.8ex] & \multicolumn{3}{c}{Log Change in Shootings} \\ 
\\[-1.8ex] & \multicolumn{1}{c}{(1)} & \multicolumn{1}{c}{(2)} & \multicolumn{1}{c}{(3)}\\ 
\hline \\[-1.8ex] 
 Cure Implemented & -0.188^{***} & -0.161^{***} & -0.204^{***} \\ 
  & (0.033) & (0.036) & (0.041) \\ 
  Years of Cure &  & -0.012 & 0.024 \\ 
  &  & (0.007) & (0.019) \\ 
  Years of Cure Sq. &  &  & -0.004^{**} \\ 
  &  &  & (0.002) \\ 
  Felony Complaints & \checkmark & \checkmark & \checkmark \\
  & & & \\ 
 Observations & \multicolumn{1}{c}{1,368} & \multicolumn{1}{c}{1,368} & \multicolumn{1}{c}{1,368} \\ 
\hline \\[-1.8ex] 
\end{tabular}

\raggedright
\small{\textit{Note}: This table replicates Table \ref{tab:mainPoisDIDsTab}, but additionally controls for the number of felony complaints reported in each precinct-year. All models include precinct and year fixed effects. Robust standard errors are in parentheses. * p<0.1, ** p<0.05, *** p<0.01.}
\end{table} 

\begin{table}[!htbp] \centering 
  \caption{Impact of Cure DID Results} 
  \label{tab:mainPoisDIDsComplaintsArrestsTab} 
\begin{tabular}{@{\extracolsep{5pt}}lD{.}{.}{-3} D{.}{.}{-3} D{.}{.}{-3} } 
\\[-1.8ex]\hline 
\hline \\[-1.8ex] 
\\[-1.8ex] & \multicolumn{3}{c}{Log Change in Shootings} \\ 
\\[-1.8ex] & \multicolumn{1}{c}{(1)} & \multicolumn{1}{c}{(2)} & \multicolumn{1}{c}{(3)}\\ 
\hline \\[-1.8ex] 
 Cure Implemented & -0.190^{***} & -0.162^{***} & -0.206^{***} \\ 
  & (0.034) & (0.036) & (0.042) \\ 
  Years of Cure &  & -0.012^{*} & 0.024 \\ 
  &  & (0.007) & (0.019) \\ 
  Years of Cure Sq. &  &  & -0.004^{**} \\ 
  &  &  & (0.002) \\ 
  Felony Complaints & \checkmark & \checkmark & \checkmark \\
  & & & \\
  Felony Arrests & \checkmark & \checkmark & \checkmark \\
  & & & \\ 
 Observations & \multicolumn{1}{c}{1,368} & \multicolumn{1}{c}{1,368} & \multicolumn{1}{c}{1,368} \\ 
\hline \\[-1.8ex] 
\end{tabular}

\raggedright
\small{\textit{Note}: This table replicates Table \ref{tab:mainPoisDIDsTab}, but additionally controls for the number of felony complaints reported and the number of felony arrests made in each precinct-year. All models include precinct and year fixed effects. Robust standard errors are in parentheses. * p<0.1, ** p<0.05, *** p<0.01.}
\end{table}

\begin{table}[!htbp] \centering 
  \caption{Impact of Cure Event Study} 
  \label{tab:mainPoisESComplaintsTab} 
\begin{tabular}{@{\extracolsep{5pt}}lD{.}{.}{-3} } 
\\[-1.8ex]\hline 
\hline \\[-1.8ex] 
\\[-1.8ex] & \multicolumn{1}{c}{Log Change in Shootings} \\ 
\hline \\[-1.8ex] 
 <=6 years pre-Cure & -0.013 \\ 
  & (0.057) \\ 
  5 years pre-Cure & -0.017 \\ 
  & (0.063) \\ 
  4 years pre-Cure & 0.001 \\ 
  & (0.060) \\ 
  3 years pre-Cure & 0.006 \\ 
  & (0.060) \\ 
  2 years pre-Cure & 0.030 \\ 
  & (0.068) \\ 
  Cure start year & -0.195^{***} \\ 
  & (0.064) \\ 
  1 year post-Cure & -0.186^{***} \\ 
  & (0.063) \\ 
  2 years post-Cure & -0.200^{**} \\ 
  & (0.078) \\ 
  >=3 years post-Cure & -0.177^{***} \\ 
  & (0.057) \\ 
  Felony Complaints & \checkmark \\
  & \\
 Observations & \multicolumn{1}{c}{1,368} \\ 
\hline \\[-1.8ex] 
\end{tabular}

\raggedright
\small{\textit{Note}: This table replicates Table \ref{tab:mainPoisESTab}, but additionally controls for the number of felony complaints reported in each precinct-year. The first year pre-Cure is the reference period. The model includes precinct and year fixed effects. Robust standard errors are in parentheses. * p<0.1, ** p<0.05, *** p<0.01.}
\end{table} 

\begin{table}[!htbp] \centering 
  \caption{Impact of Cure Event Study} 
  \label{tab:mainPoisESComplaintsArrestsTab} 
\begin{tabular}{@{\extracolsep{5pt}}lD{.}{.}{-3} } 
\\[-1.8ex]\hline 
\hline \\[-1.8ex] 
\\[-1.8ex] & \multicolumn{1}{c}{Log Change in Shootings} \\ 
\hline \\[-1.8ex] 
 <=6 years pre-Cure & -0.012 \\ 
  & (0.057) \\ 
  5 years pre-Cure & -0.017 \\ 
  & (0.063) \\ 
  4 years pre-Cure & 0.001 \\ 
  & (0.060) \\ 
  3 years pre-Cure & 0.006 \\ 
  & (0.060) \\ 
  2 years pre-Cure & 0.030 \\ 
  & (0.068) \\ 
  Cure start year & -0.195^{***} \\ 
  & (0.064) \\ 
  1 year post-Cure & -0.187^{***} \\ 
  & (0.064) \\ 
  2 years post-Cure & -0.200^{**} \\ 
  & (0.078) \\ 
  >=3 years post-Cure & -0.179^{***} \\ 
  & (0.057) \\ 
  Felony Complaints & \checkmark \\
  & \\ 
  Felony Arrests & \checkmark \\
  & \\ 
 Observations & \multicolumn{1}{c}{1,368} \\ 
\hline \\[-1.8ex] 
\end{tabular}

\raggedright
\small{\textit{Note}: This table replicates Table \ref{tab:mainPoisESTab}, but additionally controls for the number of felony complaints reported and the number of felony arrests made in each precinct-year. The first year pre-Cure is the reference period. The model includes precinct and year fixed effects. Robust standard errors are in parentheses. * p<0.1, ** p<0.05, *** p<0.01.}
\end{table}
\newpage

\begin{table}[!htbp] \centering 
  \caption{Impact of Cure: Only Cure Precincts}
  \label{tab:onlyCurePoisDIDsComplaintsTab} 
\begin{tabular}{@{\extracolsep{5pt}}lD{.}{.}{-3} D{.}{.}{-3} D{.}{.}{-3} } 
\\[-1.8ex]\hline 
\hline \\[-1.8ex] 
\\[-1.8ex] & \multicolumn{3}{c}{Log Change in Shootings} \\ 
\\[-1.8ex] & \multicolumn{1}{c}{(1)} & \multicolumn{1}{c}{(2)} & \multicolumn{1}{c}{(3)}\\ 
\hline \\[-1.8ex] 
 Cure Implemented & -0.109^{***} & -0.108^{***} & -0.143^{***} \\ 
  & (0.041) & (0.041) & (0.047) \\ 
  Years of Cure &  & -0.002 & 0.027 \\ 
  &  & (0.009) & (0.020) \\ 
  Years of Cure Sq. &  &  & -0.003 \\ 
  &  &  & (0.002) \\ 
  Felony Complaints & \checkmark & \checkmark & \checkmark \\
  & & & \\
 Observations & \multicolumn{1}{c}{504} & \multicolumn{1}{c}{504} & \multicolumn{1}{c}{504} \\ 
\hline \\[-1.8ex] 
\end{tabular}

\raggedright
\small{\textit{Note}: This table replicates Table \ref{tab:onlyCurePoisDIDsTab}, but additionally controls for the number of felony complaints reported in each precinct-year. All models include precinct and year fixed effects. Robust standard errors are in parentheses. * p<0.1, ** p<0.05, *** p<0.01.}
\end{table} 

\begin{table}[!htbp] \centering 
  \caption{Impact of Cure: Only Cure Precincts}
  \label{tab:onlyCurePoisDIDsComplaintsArrestsTab} 
\begin{tabular}{@{\extracolsep{5pt}}lD{.}{.}{-3} D{.}{.}{-3} D{.}{.}{-3} } 
\\[-1.8ex]\hline 
\hline \\[-1.8ex] 
\\[-1.8ex] & \multicolumn{3}{c}{Log Change in Shootings} \\ 
\\[-1.8ex] & \multicolumn{1}{c}{(1)} & \multicolumn{1}{c}{(2)} & \multicolumn{1}{c}{(3)}\\ 
\hline \\[-1.8ex] 
 Cure Implemented & -0.111^{***} & -0.109^{***} & -0.146^{***} \\ 
  & (0.041) & (0.041) & (0.047) \\ 
  Years of Cure &  & -0.005 & 0.026 \\ 
  &  & (0.009) & (0.020) \\ 
  Years of Cure Sq. &  &  & -0.004^{*} \\ 
  &  &  & (0.002) \\ 
  Felony Complaints & \checkmark & \checkmark & \checkmark \\
  & & & \\ 
  Felony Arrests & \checkmark & \checkmark & \checkmark \\
  & & & \\ 
 Observations & \multicolumn{1}{c}{504} & \multicolumn{1}{c}{504} & \multicolumn{1}{c}{504} \\ 
\hline \\[-1.8ex] 
\end{tabular} 

\raggedright
\small{\textit{Note}: This table replicates Table \ref{tab:onlyCurePoisDIDsTab}, but additionally controls for the number of felony complaints reported and the number of felony arrests made in each precinct-year. All models include precinct and year fixed effects. Robust standard errors are in parentheses. * p<0.1, ** p<0.05, *** p<0.01.}
\end{table} 

\begin{table}[!htbp] \centering 
  \caption{Spillover Effects}
  \label{tab:SOBeforeCurePoisSODIDsComplaintsTab} 
\begin{tabular}{@{\extracolsep{5pt}}lD{.}{.}{-3} D{.}{.}{-3} D{.}{.}{-3} } 
\\[-1.8ex]\hline 
\hline \\[-1.8ex] 
\\[-1.8ex] & \multicolumn{3}{c}{Log Change in Shootings} \\ 
\\[-1.8ex] & \multicolumn{1}{c}{(1)} & \multicolumn{1}{c}{(2)} & \multicolumn{1}{c}{(3)}\\ 
\hline \\[-1.8ex] 
 Spillover Possible & -0.127^{***} & -0.038 & -0.113^{**} \\ 
  & (0.046) & (0.047) & (0.054) \\ 
  Years of Spillover &  & -0.048^{***} & 0.019 \\ 
  &  & (0.009) & (0.023) \\ 
  Years of Spillover Sq. &  &  & -0.007^{***} \\ 
  &  &  & (0.002) \\ 
  Felony Complaints & \checkmark & \checkmark & \checkmark \\
  & & & \\ 
 Observations & \multicolumn{1}{c}{1,134} & \multicolumn{1}{c}{1,134} & \multicolumn{1}{c}{1,134} \\ 
\hline \\[-1.8ex] 
\end{tabular}

\raggedright
\small{\textit{Note}: This table replicates Table \ref{tab:SOBeforeCurePoisSODIDsTab}, but additionally controls for the number of felony complaints reported in each precinct-year. All models include precinct and year fixed effects. Robust standard errors are in parentheses. * p<0.1, ** p<0.05, *** p<0.01.}
\end{table}

\begin{table}[!htbp] \centering 
  \caption{Spillover Effects}
  \label{tab:SOBeforeCurePoisSODIDsComplaintsArrestsTab} 
\begin{tabular}{@{\extracolsep{5pt}}lD{.}{.}{-3} D{.}{.}{-3} D{.}{.}{-3} } 
\\[-1.8ex]\hline 
\hline \\[-1.8ex] 
\\[-1.8ex] & \multicolumn{3}{c}{Log Change in Shootings} \\ 
\\[-1.8ex] & \multicolumn{1}{c}{(1)} & \multicolumn{1}{c}{(2)} & \multicolumn{1}{c}{(3)}\\ 
\hline \\[-1.8ex] 
 Spillover Possible & -0.125^{***} & -0.038 & -0.113^{**} \\ 
  & (0.047) & (0.047) & (0.054) \\ 
  Years of Spillover &  & -0.048^{***} & 0.020 \\ 
  &  & (0.009) & (0.024) \\ 
  Years of Spillover Sq. &  &  & -0.007^{***} \\ 
  &  &  & (0.002) \\ 
  Felony Complaints & \checkmark & \checkmark & \checkmark \\
  & & & \\ 
  Felony Arrests & \checkmark & \checkmark & \checkmark \\
  & & & \\ 
 Observations & \multicolumn{1}{c}{1,134} & \multicolumn{1}{c}{1,134} & \multicolumn{1}{c}{1,134} \\ 
\hline \\[-1.8ex] 
\end{tabular}

\raggedright
\small{\textit{Note}: This table replicates Table \ref{tab:SOBeforeCurePoisSODIDsTab}, but additionally controls for the number of felony complaints reported and the number of felony arrests made in each precinct-year. All models include precinct and year fixed effects. Robust standard errors are in parentheses. * p<0.1, ** p<0.05, *** p<0.01.}
\end{table}

\begin{table}[!htbp] \centering 
  \caption{Spillover Effects: Only Control Precincts}
  \label{tab:onlyControlsPoisSODIDsComplaintsTab} 
\begin{tabular}{@{\extracolsep{5pt}}lD{.}{.}{-3} D{.}{.}{-3} D{.}{.}{-3} } 
\\[-1.8ex]\hline 
\hline \\[-1.8ex] 
\\[-1.8ex] & \multicolumn{3}{c}{Log Change in Shootings} \\ 
\\[-1.8ex] & \multicolumn{1}{c}{(1)} & \multicolumn{1}{c}{(2)} & \multicolumn{1}{c}{(3)}\\ 
\hline \\[-1.8ex] 
 Spillover Possible & -0.160^{**} & -0.090 & -0.124 \\ 
  & (0.064) & (0.070) & (0.085) \\ 
  Years of Spillover &  & -0.026^{*} & 0.000 \\ 
  &  & (0.013) & (0.036) \\ 
  Years of Spillover Sq. &  &  & -0.003 \\ 
  &  &  & (0.004) \\ 
  Felony Complaints & \checkmark & \checkmark & \checkmark \\
  & & & \\ 
 Observations & \multicolumn{1}{c}{864} & \multicolumn{1}{c}{864} & \multicolumn{1}{c}{864} \\ 
\hline \\[-1.8ex] 
\end{tabular}

\raggedright
\small{\textit{Note}: This table replicates Table \ref{tab:onlyControlsPoisSODIDsTab}, but additionally controls for the number of felony complaints reported in each precinct-year. All models include precinct and year fixed effects. Robust standard errors are in parentheses. * p<0.1, ** p<0.05, *** p<0.01.}
\end{table} 

\begin{table}[!htbp] \centering 
  \caption{Spillover Effects: Only Control Precincts}
  \label{tab:onlyControlsPoisSODIDsComplaintsArrestsTab} 
\begin{tabular}{@{\extracolsep{5pt}}lD{.}{.}{-3} D{.}{.}{-3} D{.}{.}{-3} } 
\\[-1.8ex]\hline 
\hline \\[-1.8ex] 
\\[-1.8ex] & \multicolumn{3}{c}{Log Change in Shootings} \\ 
\\[-1.8ex] & \multicolumn{1}{c}{(1)} & \multicolumn{1}{c}{(2)} & \multicolumn{1}{c}{(3)}\\ 
\hline \\[-1.8ex] 
 Spillover Possible & -0.138^{**} & -0.090 & -0.115 \\ 
  & (0.066) & (0.071) & (0.086) \\ 
  Years of Spillover &  & -0.019 & -0.000 \\ 
  &  & (0.014) & (0.036) \\ 
  Years of Spillover Sq. &  &  & -0.002 \\ 
  &  &  & (0.004) \\ 
  Felony Complaints & \checkmark & \checkmark & \checkmark \\
  & & & \\ 
  Felony Arrests & \checkmark & \checkmark & \checkmark \\
  & & & \\
 Observations & \multicolumn{1}{c}{864} & \multicolumn{1}{c}{864} & \multicolumn{1}{c}{864} \\ 
\hline \\[-1.8ex] 
\end{tabular}

\raggedright
\small{\textit{Note}: This table replicates Table \ref{tab:onlyControlsPoisSODIDsTab}, but additionally controls for the number of felony complaints reported and the number of felony arrests made in each precinct-year. All models include precinct and year fixed effects. Robust standard errors are in parentheses. * p<0.1, ** p<0.05, *** p<0.01.}
\end{table}

\begin{table}[!htbp] \centering 
  \caption{Impact of Cure on Post-Spillover Precincts}
  \label{tab:NonSOControls_SOBeforeCurePoisDIDsComplaintsTab} 
\begin{tabular}{@{\extracolsep{5pt}}lD{.}{.}{-3} D{.}{.}{-3} D{.}{.}{-3} } 
\\[-1.8ex]\hline 
\hline \\[-1.8ex] 
\\[-1.8ex] & \multicolumn{3}{c}{Log Change in Shootings} \\ 
\\[-1.8ex] & \multicolumn{1}{c}{(1)} & \multicolumn{1}{c}{(2)} & \multicolumn{1}{c}{(3)}\\ 
\hline \\[-1.8ex] 
 Cure Implemented & -0.208^{***} & -0.180^{***} & -0.320^{***} \\ 
  & (0.055) & (0.058) & (0.066) \\ 
  Years of Cure &  & -0.016 & 0.124^{***} \\ 
  &  & (0.013) & (0.038) \\ 
  Years of Cure Sq. &  &  & -0.019^{***} \\ 
  &  &  & (0.005) \\ 
  Felony Complaints & \checkmark & \checkmark & \checkmark \\
  & & & \\ 
 Observations & \multicolumn{1}{c}{630} & \multicolumn{1}{c}{630} & \multicolumn{1}{c}{630} \\ 
\hline \\[-1.8ex] 
\end{tabular}

\raggedright
\small{\textit{Note}: This table replicates Table \ref{tab:NonSOControls_SOBeforeCurePoisDIDsTab}, but additionally controls for the number of felony complaints reported in each precinct-year. All models include precinct and year fixed effects. Robust standard errors are in parentheses. * p<0.1, ** p<0.05, *** p<0.01.}
\end{table} 

\begin{table}[!htbp] \centering 
  \caption{Impact of Cure on Post-Spillover Precincts}
  \label{tab:NonSOControls_SOBeforeCurePoisDIDsComplaintsArrestsTab} 
\begin{tabular}{@{\extracolsep{5pt}}lD{.}{.}{-3} D{.}{.}{-3} D{.}{.}{-3} } 
\\[-1.8ex]\hline 
\hline \\[-1.8ex] 
\\[-1.8ex] & \multicolumn{3}{c}{Log Change in Shootings} \\ 
\\[-1.8ex] & \multicolumn{1}{c}{(1)} & \multicolumn{1}{c}{(2)} & \multicolumn{1}{c}{(3)}\\ 
\hline \\[-1.8ex] 
 Cure Implemented & -0.234^{***} & -0.202^{***} & -0.326^{***} \\ 
  & (0.058) & (0.059) & (0.066) \\ 
  Years of Cure &  & -0.020 & 0.110^{***} \\ 
  &  & (0.013) & (0.036) \\ 
  Years of Cure Sq. &  &  & -0.018^{***} \\ 
  &  &  & (0.005) \\ 
  Felony Complaints & \checkmark & \checkmark & \checkmark \\
  & & & \\ 
  Felony Arrests & \checkmark & \checkmark & \checkmark \\
  & & & \\
 Observations & \multicolumn{1}{c}{630} & \multicolumn{1}{c}{630} & \multicolumn{1}{c}{630} \\ 
\hline \\[-1.8ex] 
\end{tabular} 

\raggedright
\small{\textit{Note}: This table replicates Table \ref{tab:NonSOControls_SOBeforeCurePoisDIDsTab}, but additionally controls for the number of felony complaints reported and the number of felony arrests made in each precinct-year. All models include precinct and year fixed effects. Robust standard errors are in parentheses. * p<0.1, ** p<0.05, *** p<0.01.}
\end{table}

\begin{table}[!htbp] \centering 
  \caption{Spillover Effects on Already Treated Precincts}
  \label{tab:NonSOControls_SOAfterCurePoisSODIDsComplaintsTab} 
\begin{tabular}{@{\extracolsep{5pt}}lD{.}{.}{-3} D{.}{.}{-3} D{.}{.}{-3} } 
\\[-1.8ex]\hline 
\hline \\[-1.8ex] 
\\[-1.8ex] & \multicolumn{3}{c}{Log Change in Shootings} \\ 
\\[-1.8ex] & \multicolumn{1}{c}{(1)} & \multicolumn{1}{c}{(2)} & \multicolumn{1}{c}{(3)}\\ 
\hline \\[-1.8ex] 
 Spillover Possible & -0.290^{***} & -0.235^{***} & -0.303^{***} \\ 
  & (0.070) & (0.075) & (0.092) \\ 
  Years of Spillover &  & -0.025^{*} & 0.040 \\ 
  &  & (0.015) & (0.049) \\ 
  Years of Spillover Sq. &  &  & -0.009 \\ 
  &  &  & (0.006) \\ 
  Felony Complaints & \checkmark & \checkmark & \checkmark \\
  & & & \\
 Observations & \multicolumn{1}{c}{486} & \multicolumn{1}{c}{486} & \multicolumn{1}{c}{486} \\ 
\hline \\[-1.8ex] 
\end{tabular}

\raggedright
\small{\textit{Note}: This table replicates Table \ref{tab:NonSOControls_SOAfterCurePoisSODIDsTab}, but additionally controls for the number of felony complaints reported in each precinct-year. All models include precinct and year fixed effects. Robust standard errors are in parentheses. * p<0.1, ** p<0.05, *** p<0.01.}
\end{table}

\begin{table}[!htbp] \centering 
  \caption{Spillover Effects on Already Treated Precincts}
  \label{tab:NonSOControls_SOAfterCurePoisSODIDsComplaintsArrestsTab}  
\begin{tabular}{@{\extracolsep{5pt}}lD{.}{.}{-3} D{.}{.}{-3} D{.}{.}{-3} } 
\\[-1.8ex]\hline 
\hline \\[-1.8ex] 
\\[-1.8ex] & \multicolumn{3}{c}{Log Change in Shootings} \\ 
\\[-1.8ex] & \multicolumn{1}{c}{(1)} & \multicolumn{1}{c}{(2)} & \multicolumn{1}{c}{(3)}\\ 
\hline \\[-1.8ex] 
 Spillover Possible & -0.293^{***} & -0.242^{***} & -0.300^{***} \\ 
  & (0.075) & (0.077) & (0.092) \\ 
  Years of Spillover &  & -0.033^{**} & 0.028 \\ 
  &  & (0.016) & (0.051) \\ 
  Years of Spillover Sq. &  &  & -0.008 \\ 
  &  &  & (0.006) \\ 
  Felony Complaints & \checkmark & \checkmark & \checkmark \\
  & & & \\ 
  Felony Arrests & \checkmark & \checkmark & \checkmark \\
  & & & \\
 Observations & \multicolumn{1}{c}{486} & \multicolumn{1}{c}{486} & \multicolumn{1}{c}{486} \\ 
\hline \\[-1.8ex] 
\end{tabular}

\raggedright
\small{\textit{Note}: This table replicates Table \ref{tab:NonSOControls_SOAfterCurePoisSODIDsTab}, but additionally controls for the number of felony complaints reported and the number of felony arrests made in each precinct-year. All models include precinct and year fixed effects. Robust standard errors are in parentheses. * p<0.1, ** p<0.05, *** p<0.01.}
\end{table} 

\end{document}